\documentclass[aps,prb,twocolumn,10pt,showpacs,amsmath,amssymb,floatfix]{revtex4-1}
\usepackage{graphicx}
\usepackage{amsmath}
\usepackage{amssymb}
\usepackage[utf8]{inputenc}
\usepackage{comment}
\usepackage{dcolumn}
\usepackage{bm}
\usepackage{hyperref}
\usepackage{soul}

\newcommand{\Rmnum}[1]{\expandafter\@slowromancap\romannumeral  #1@}

\newcommand{\psih}{\hat{\psi}}
\newcommand{\psihd}{\hat{\psi}^\dagger}

\newcommand{\ch}{\hat{c}}
\newcommand{\chd}{\hat{c}^\dagger}

\newcommand{\Hh}{\hat{H}}
\newcommand{\Fh}{\hat{F}}
\newcommand{\Gh}{\hat{G}}

\newcommand{\nh}{\hat{n}}

\newcommand{\hh}{\hat{h}}

\newcommand{\Tr}{\text{Tr}}
\newcommand{\commutator}[1]{\left [ #1 \right ] }

\newcommand{\parenth}[1]{\left ( #1 \right ) }
\newcommand{\trace}[1]{\Tr\left \{ #1 \right \}}
\newcommand{\traceg}[1]{\Tr_\gamma\left \{ #1 \right \}}

\newcommand{\Eq}[1]{Eq.~\!\!(\ref{#1})}
\newcommand{\zb}{\bar{z}}

\newcommand{\Sigmah}{ {\hat{\Sigma}} }

\newcommand{\SigmaRt}{ {\tilde{\Sigma}^R} }

\newcommand{\Fig}[1]{FIG. \ref{#1}}
\newcommand{\bp}{ {\bm{p}} }

\newcommand{\lambdat}{\tilde{\lambda}}
\newcommand{\Gammah}{\hat{\Gamma}}
\newcommand{\Zh}{\hat{Z}}
\newcommand{\Jh}{\hat{J}}
\newcommand{\Kh}{\hat{J}}
\newcommand{\Ah}{\hat{A}}
\newcommand{\Qh}{\hat{Q}}
\newcommand{\traceC}[1]{\Tr_C \left \{ #1 \right \}}
\newcommand{\xb}{{\bf x}}
\renewcommand\Re{\operatorname{Re}}
\renewcommand\Im{\operatorname{Im}}
\usepackage{bbold}
\newcommand{\one}{\mathbb{1}} 
\newcommand{\oneh}{\hat{\mathbb{1}}} 

\graphicspath{{pics/}}
\begin{document}

\title{Partial self-consistency and analyticity in many-body perturbation theory: \\ particle number conservation and a generalized sum rule}

\author{Daniel Karlsson} 
\affiliation{Department of Physics,
Nanoscience Center P.O.Box 35 FI-40014 University of Jyv\"{a}skyl\"{a}, Finland}
\affiliation{European Theoretical Spectroscopy Facility, ETSF}

\author{Robert van Leeuwen}
\affiliation{Department of Physics,
Nanoscience Center P.O.Box 35 FI-40014 University of Jyv\"{a}skyl\"{a}, Finland}
\affiliation{European Theoretical Spectroscopy Facility, ETSF}

\begin{abstract}
We consider a general class of approximations which guarantees the conservation of particle number  in many-body perturbation theory. To do this we extend the concept of $\Phi$-derivability for the self-energy $\Sigma$ to a larger class of diagrammatic terms in which only some of the Green's function lines contain the fully dressed Green's function $G$.  We call the corresponding approximations for $\Sigma$ partially $\Phi$-derivable. A special subclass of such approximations, which are gauge-invariant, is obtained by dressing loops in the diagrammatic expansion of $\Phi$ consistently with $G$. These approximations are number conserving but do not have to fulfill other conservation laws, such as the conservation of energy and momentum. From our formalism we can easily deduce if commonly used approximations will fulfill the continuity equation, which implies particle number conservation. We further show how the concept of partial $\Phi$-derivability plays an important role in the derivation of a generalized sum rule for the particle number, which reduces to the Luttinger-Ward theorem in the case of a homogeneous electron gas, and the Friedel sum rule in the case of the Anderson model. To do this we need to ensure that the Green's function has certain complex analytic properties, which can be guaranteed if the spectral function is positive semi-definite. The latter property can be ensured for a subset of partially $\Phi$-derivable approximations for the self-energy, namely those that can be constructed from squares of so-called half-diagrams. In case the analytic requirements are not fulfilled we highlight a number of subtle issues related to branch cuts, pole structure and multi-valuedness. We also show that various schemes of computing the particle number are consistent for particle number conserving approximations.
\end{abstract}


\maketitle
\section{Introduction}

Many-body perturbation theory using Green's functions is a successful and powerful method for studying interacting systems in and out of equilibrium. Its most general non-equilibrium version~\cite{stefanucci2013} is routinely used for very diverse situations such as quantum transport (e.g. through single molecules~\cite{Thygesen2008}), cold atoms in optical lattices~\cite{Schlunzen2016} and transient dynamics (e.g. transient photo-absorption\cite{Perfetto2015b}). In the limiting case  of equilibrium systems it can also be used to study standard photo-absorption and photo-emission spectroscopies~\cite{Strinati1988,Bechstedt2015}. Moreover the non-equilibrium formalism even in this limiting case has found useful applications to derive approximations that guarantee positive spectral distributions as it allows for a direct expansion for those quantities~\cite{Stefanucci2014,StefanucciErratum,Uimonen2015}.

One of the reasons of the successes of many-body perturbation theory is the use of diagrammatic techniques, where one can sum up diagrams to infinite order in a practical way. However, if partial resummations are used, when one sums over a subset over all diagrams, important conservation laws can be violated. The most important laws are the fulfillment of the continuity equation, momentum and angular momentum conservation, and energy conservation. Approximations which fulfill all of these conservation laws are referred to as conserving~\cite{Baym1961}. 

Baym~\cite{Baym1962} considered approximate generating functionals for the single-particle Green's function $G$, the $\Phi$-functional, from which the self-energy $\Sigma$ could be constructed via $\Sigma = \delta \Phi / \delta G$. If $G$ is obtained from the self-consistent solution of Dyson's equation for such a $\Sigma$, the approximation is called $\Phi$-derivable and is automatically conserving. Commonly used $\Phi$-derivable many-body approximations (fully self-consistent) are the Hartree-Fock approximation, the Second Born Approximation (2BA), the $GW$ approximation and the T-matrix approximation.

While fully self-consistent schemes have many conceptual advantages, they usually carry a high computation cost, as well as other features that are not desirable, both in and out of equilibrium. In fully self-consistent calculations, it has been shown \cite{Friesen2009, PuigvonFriesen2010} that strong time-dependent fields can yield artificial steady states in finite systems. Partial self-consistency lessened this effect. Another method related to partial self-consistency is the Generalized Kadanoff-Baym Ansatz (GKBA), where one approximates certain time-nondiagonal elements of the Green's function and can be made conserving, which is also free of artificial damping \cite{Hermanns2014a,Kremp1999,Kremp2005}.

In equilibrium, especially in the \emph{ab-initio }community, the vast majority of calculations are non-selfconsistent~\cite{Bechstedt2015}, or has a small degree of partial self-consistency in it \cite{VanSchilfgaarde2006,Kotani2007} (However, see Refs.~\cite{DeGroot1995,Holm1998,Stan2006,Rostgaard2010,Caruso2012a,Caruso2013a,Koval2014} for examples of fully self-consistent $GW$ calculations).  These types of approximations tend to yield accurate band gaps and
spectral properties.  Furthermore, for $GW$ for the electron gas, full self-consistency worsens the spectral features compared to non-selfconsistent calculations\cite{VonBarth1996,Holm1998}, even if the scheme is energy conserving. While these drawbacks will be cured by vertex corrections to $GW$, there has been a long discussion in the literature if one should focus on non-selfconsistent vertex corrections, or full self-consistency (see, for example, \cite{Aryasetiawan1998,Aulbur2000,Aryasetiawan2009}).
Moreover, partially self-consistent approximations are expected to yield reasonable energies, especially if one uses the Luttinger-Ward functional for calculating the total energy \cite{Almbladh1999}.

Thus, while the concept of a $\Phi$-derivable approximation is extremely useful, it is quite restrictive. As soon as a scheme is not fully self-consistent, the concept of $\Phi$-derivability is lost, together with a convenient way of seeing if the chosen approximation is particle number conserving. The class of approximations that do conserve particle number is larger than those generated by $\Phi$-functionals, however. For example, the partially self-consistent $GW_0$ approximation (where the screened interaction $W$ is kept fixed to be $W_0$ during the self-consistency cycle)  and the non-selfconsistent $G_0W_0$ approximation are not $\Phi$-derivable. However, $GW_0$ is particle number conserving (first shown for the equilibrium electron gas\cite{Holm1997, VonBarth1996}, and later more generally\cite{Schindlmayr2001}) but not fully conserving~\cite{Stan2009a}, while $G_0W_0$ is not particle number conserving~\cite{Schindlmayr1997,Schindlmayr2001}.

Another property that can be violated by summing over subclasses of diagrams is the positivity of the spectral functions of the Green's function and the polarizability. Positivity is an important ingredient for two reasons. First of all, it is related to a probability interpretation of photo-emission and photo-absorption processes, which naturally requires positive probabilities. Second, it guarantees correct analyticity and causality properties of the various propagators, which is a necessary condition for performing self-consistent calculations, as the analytic properties in general will deteriorate every iteration cycle.

The conserving and analytic properties are related but not equivalent. A situation in which both of these properties play a role is in the derivation of sum rules for the particle number, such as the Luttinger-Ward~\cite{Luttinger1960a} for the electron gas and the Friedel sum rule~\cite{Langreth1966,Langer1961} for the Anderson model. Another occasion is the derivation of the Ward identity and the frequency sum rule for the density response function. Conserving approximation in connection with the Luttinger-Ward sum rule were also studied in the context of the non-perturbative approximation of dynamical mean field theory in Ref.~\cite{Ortloff2007}.

In this work we generalize the concept of $\Phi$-derivability to partially self-consistent schemes. While the resulting approximations are not fully conserving, a large class of many-body approximations used in the literature is shown to be included in this larger class of approximations. The formalism provides an easy way to see if an approximation conserves particle number or not. Since we are using NEGF, the concept of partially self-consistent $\Phi$-functionals is valid both in and out of equilibrium. Using the new formalism, we study particle number sum rules in equilibrium in their most general settings, and highlight the importance of particle conservation. We state sufficient conditions for the sum rules to be valid in a given diagrammatic approximation. In particular, we stress that the sum rules are valid provided that $G$ has the correct analytical properties, and we study the consequences of incorrect analytical properties, including issues related to pole structure and multivaluedness of the logarithm. We exemplify the formalism and sum rules in model systems of quantum transport.

The paper is structured as follows: In the first part of this work, Sec.\ref{phiDerivability},  we will introduce the concept of partial $\Phi$-derivability. Subsequently in Sec.~\ref{PSD} we will discuss the complex analytic properties for approximate Green's function and self-energies and the way these properties affect the calculation of the particle number. Then finally, in Sec.~\ref{sumRules}, we discuss the derivation of a generalized sum rule for the particle number and illustrate the concepts by explicit numerical calculations. We end with our conclusions in Sec.~\ref{conclusions}.

\section{Partial \texorpdfstring{$\Phi$}{Phi}-derivability}\label{phiDerivability}

Our main object of study is the single-particle Green's function $G(1,2)$, where $1 = ({\bf r}_1, \sigma_1, z_1)$ is a collective space-spin-time variable. 
The Green's function is defined on a time contour $\gamma$ that has a forward, backward, and Matsubara (imaginary-time) branch,~\cite{stefanucci2013} 
\begin{align}
 G(1,2) = (-i) \frac{\trace{ \mathcal{T} \left [ e^{-i\int_\gamma d\zb \Hh(\zb)} \psih (1) \psih^\dagger(2) \right ]   }} 
 {\trace{ \mathcal{T} \left[ e^{-i\int_\gamma d\zb \Hh(\zb)} \right ]}},
\end{align}
where $\psih$ $(\psihd)$ are fermionic field destruction (creation) operators, and $\mathcal{T}$ is the time-ordering operator on the contour $\gamma$. The time-dependent Hamiltonian  $\Hh(z) = \Hh_0(z) + \Hh_{\text{int}}(z)$ is given by
\begin{align}
 \Hh_0(z)             &= \int d\xb \       \psihd(\xb) h(\xb,z) \psih(\xb)   \\ 
 \Hh_{\text{int}} (z) &= \frac{1}{2}\int d\xb d\xb'\ \psihd(\xb) \psihd(\xb') v(\xb,\xb',z) \psih(\xb') \psih(\xb),
\end{align}
where $h(\xb,z)$ is the single-particle part and $v(\xb,\xb',z)$ is the two-body interaction. The non-equilibrium formalism allows for both time-dependent single-particle potentials, such as electric fields, as well as time-dependent two-body interactions, such as an adiabatic switch-on. To keep the discussion general, the time-dependence will not be specified further at this point.  We also define, for convenience, $\xb = ({\bf r},\sigma)$ and
\begin{align}
 \Gh(z_1,z_2) = \int d\xb_1 d\xb_2 \ | \xb_1 \rangle G(1,2) \langle \xb_2 |
\end{align}
to highlight the temporal dependence without being dependent on the spatial basis.

The reason for considering the contour-ordered $G$ is that general time-dependent systems at finite temperature can be considered, in and out of equilibrium. When the contour times $z$ are on the Matsubara branch, we obtain the Matsubara Green's function as $\Gh^M(\tau_1,\tau_2) = \Gh(t_0 -i\tau_1, t_0-i\tau_2)$. For general time-dependent systems, we obtain the lesser $\Gh^<$ and greater $\Gh^>$ Green's functions from $\Gh^< (t_1,t_2) = \Gh(t_1^-,t_2^+)$, and $\Gh^>(t_1,t_2) = \Gh(t_1^+,t_2^-)$, where $z=t^- / t^+$ is on the forward/backward branch. All single-particle quantities can 
be obtained by the contour-ordered Green's function.  For example, if we take $t_1=t_2$, we obtain the time-dependent density $n(\xb,t)$ and the time-dependent current density ${\bm j}(\xb,t)$  as 
\begin{align}
n(\xb,t) &= -i G^<(\xb,t;\xb,t) \\
{\bm j}(\xb,t) &= - \left [\frac{\nabla - \nabla'}{2m} G^<(\xb,t;\xb',t)\right ]_{\xb'=\xb}.
\end{align}

The equations of motion for $G(1,2)$ is given by the Kadanoff-Baym equations~\cite{Kadanoff1962,stefanucci2013},
\begin{align}
\begin{split} 
 \left [ i\frac{d}{dz_1} - \hh(z_1) \right ] \Gh(z_1,z_2) = \delta(z_1,z_2) + \\
+ \int _\gamma dz_3 \ \Sigmah(z_1,z_3) \Gh(z_3,z_2) \label{KBE1}
\end{split}
\end{align}
\begin{align}
\begin{split} 
\left [ -i\frac{d}{dz_2} - \hh(z_2) \right ] \Gh(z_1,z_2) = \delta(z_1,z_2) + \\
+ \int _\gamma dz_3 \ \Gh(z_1,z_3) \Sigmah (z_3,z_2) \label{KBE2}
\end{split}
\end{align}
with the Kubo-Martin-Schwinger (KMS) boundary conditions $\Gh(t_0-i\beta,z_2) = - \Gh(t_{0},z_2)$ and $\Gh(z_1,t_0-i\beta) = - \Gh(z_1,t_{0})$. $\Sigmah(z_1,z_2)$ is the self-energy, which has to be approximated. 
An equivalent way of writing the equations of motion is given by the Dyson equations, 
\begin{align}
\begin{split}
 G(1,2) &= \tilde{G}_0(1,2) + \! \! \iint_\gamma \!\! d3 d4 \  \tilde{G}_0(1,3) \Sigma (3,4) G(4,2) \\
 G(1,2) &= \tilde{G}_0(1,2) + \! \! \iint_\gamma \!\! d3 d4 \  G(1,3) \Sigma (3,4) \tilde{G}_0(4,2),
 \end{split}
 \label{Dyson}
\end{align}
where $\tilde{G}_0(1,2)$ is the corresponding Green's function with $\Hh_{\text{int}}=0$ that satisfies the KMS boundary conditions. 

In a given approximation to $\Sigmah$, diagrammatic or nor, there is no guarantee that the resulting scheme will be conserving. Since we in this work will focus on the fulfillment of the continuity equation, we here briefly describe it. We subtract \Eq{KBE2} from \Eq{KBE1}, and put $2=1^+$, where $1^+$ denotes $(\xb_1,z_1^+)$ and $z_1^+$ denotes a time infinitesimally later than $z_1$ on the contour. We then obtain
\begin{align}
\begin{split}
 \frac{\partial}{\partial z_1}  n (1) + \nabla \cdot \bm{j}(1) =\\
 = \int d3 \commutator{ \Sigma (1,3)G(3,1^+) - G(1,3)\Sigma (3,1^+)}.
 \label{continuityEquation}
\end{split}
 \end{align}

\Eq{continuityEquation} is a continuity equation on the contour, with a source/drain term caused by the interactions. Thus, approximations that guarantee (local) particle conservation fulfill 
\begin{align}
\int d3 \commutator{ \Sigma (1,3)G(3,1^+) - G(1,3)\Sigma (3,1^+)} = 0
\end{align}
 for all times, which is then a condition on the approximate $\Sigma(1,2)$. 
If this condition is fulfilled, the total particle number
\begin{equation}
N = - i \int d\xb  \, G^<(\xb,t;\xb,t)
\label{N_particle}
\end{equation}
is conserved in time. 

Particle number conservation is also relevant in equilibrium situations, as can be seen as follows. 
Let us consider a finite system in which the particles are non-interacting at $t_0$ with a switch-on of the interaction for times $t>t_0$. If the continuity equation is fulfilled for all times, the switching of the interaction cannot change the number of particles. In particular, this is true for an adiabatic switch on of the interaction. Thus, particle conservation is an issue also in equilibrium calculations, a point also stressed in Refs.~\cite{Schindlmayr2001,Stan2009a}. For example, within a conserving approximation the particle number can not
depend on the bond length of a molecule~\cite{Stan2009a}.

Baym \cite{Baym1962} introduced a convenient way to generate approximations that are automatically conserving, with the use of a diagrammatically defined functional $\Phi[G]$. If $\Phi[G]$ is invariant under gauge transformations, translations, rotations and time, the self-energy given via 
\begin{align}
 \Sigma [G] (1,2) = \frac{\delta \Phi [G]}{\delta G(2,1^+)}, \label{self-energy}
\end{align}
will yield a conserving approximation when the Kadanoff-Baym equations, \Eq{KBE1} and \Eq{KBE2}, are solved self-consistently with this $\Sigma$. In particular, the source term in the continuity equation, \Eq{continuityEquation}, will vanish. The scheme is then said to be $\Phi$-derivable. 

\begin{figure}
 \includegraphics[width=0.48\textwidth]{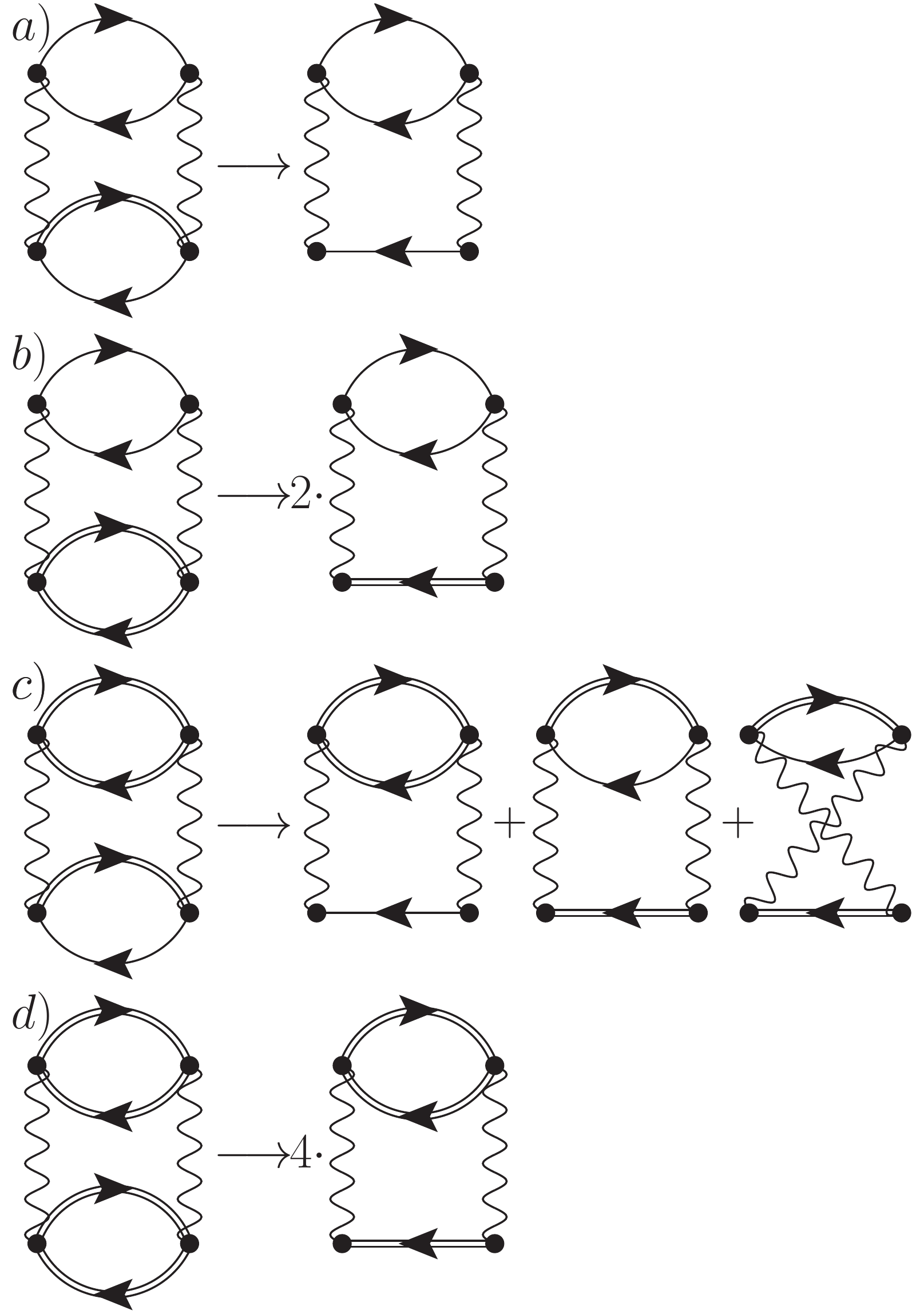}
 \caption{Examples of partially dressed ring $\Phi$-functionals (left) and the corresponding $\Sigma = \frac{\delta \Phi}{\delta G}$ (right). 
 a) A single partially dressed loop, thus $\Phi$ is not gauge invariant and the approximation is not number conserving. This diagram is part of one-shot 2nd Born, as well as $G_0W_0$. 
 b) A single, fully dressed loop, which gives a gauge-invariant, and thus particle-conserving approximation. This diagram is part of the $GW_0$ approximation. 
 c) Partially dressed loops, not gauge invariant. Several types of self-energy diagrams are obtained, where the first is part of the $G_0W$ approximation. 
 d) Fully dressed, thus conserving. Part of 2nd Born, and $GW$.
}
 \label{fig:ringDiagrams}
\end{figure}

\begin{figure}
 \includegraphics[width=0.48\textwidth]{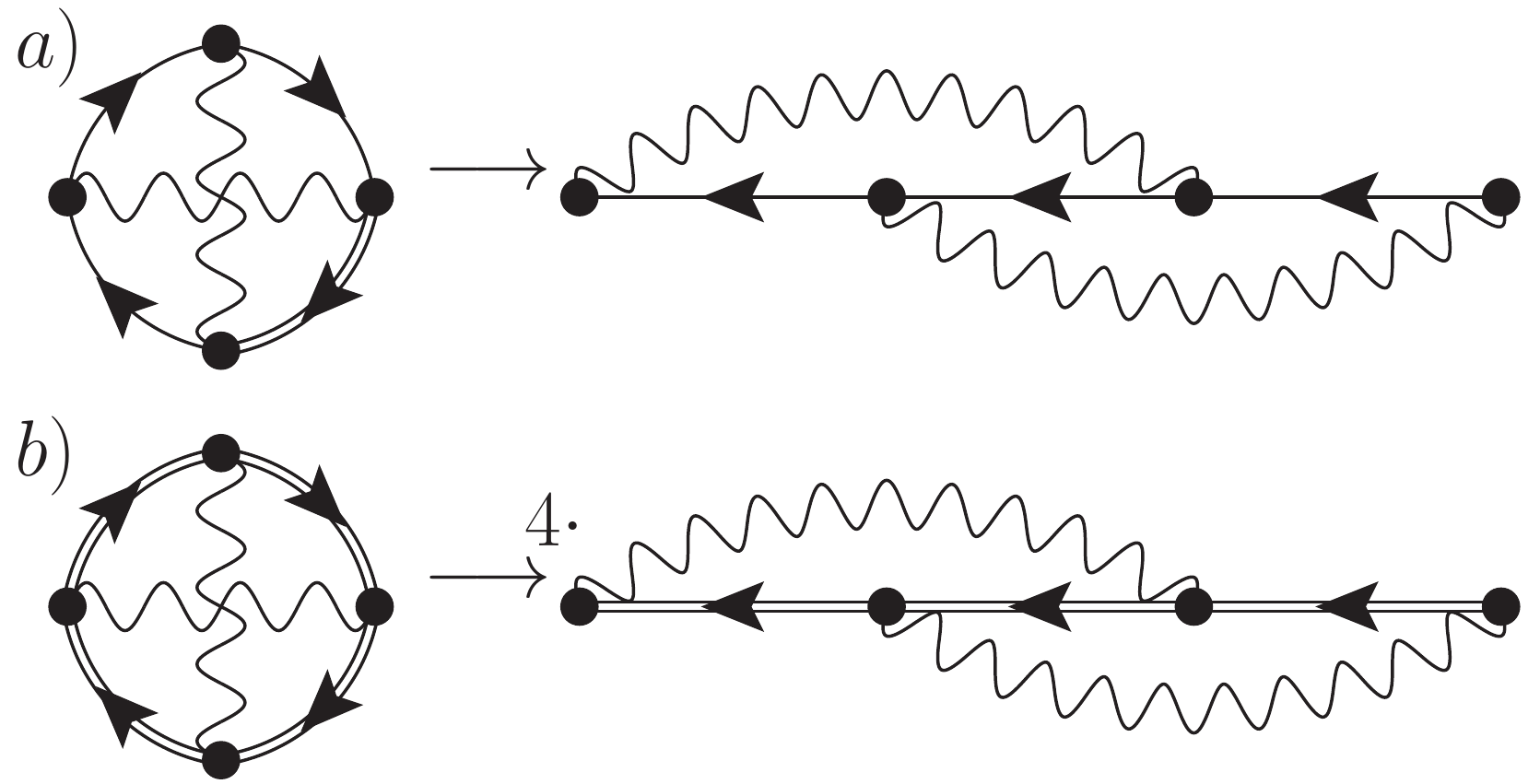}
 \caption{Examples of partially dressed 2nd Order Exchange (2OE) $\Phi$-functionals (left) and the corresponding $\Sigma = \frac{\delta \Phi}{\delta G}$ (right). a) Partially dressed second-order exchange, not gauge invariant. Constitutes one-shot 2nd Born, together with \Fig{fig:ringDiagrams}a).
 b) Fully dressed second-order exchange, thus conserving. Constitutes 2nd Born, together with \Fig{fig:ringDiagrams}d). The only way to obtain a number conserving approximation for these types of diagrams is to fully dress all $G$-lines.  
 }
 \label{fig:2ndOrderExchange}
\end{figure}

\begin{figure}
 \centering
 \includegraphics[width=0.48\textwidth]{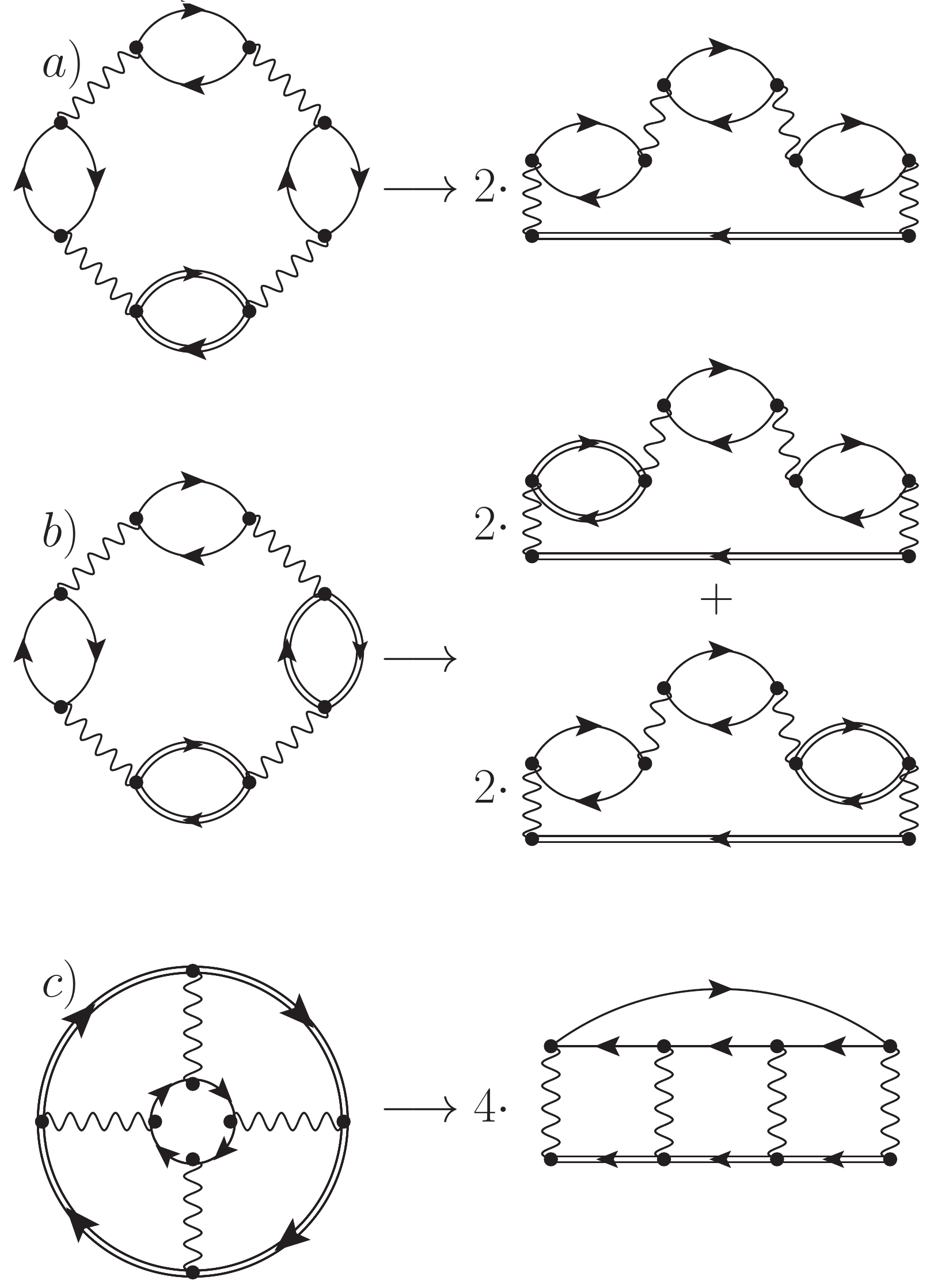}
 \caption{Examples of partially dressed, gauge invariant, $\Phi$-functionals (left) which fulfill particle number conservation, and their corresponding $\Sigma = \frac{\delta \Phi}{\delta G}$ (right). 
 a) A 4th order $GW_0$ diagram.
 b) Dressing two loops yields two non-equivalent classes of diagrams.
 c) A 4th order diagram in the particle-particle T-matrix approximation. Note that we would get the same diagram if we dress only the inner loop. 
 }
 \label{fig:fulfillingNumberConservation}
\end{figure}

We now extend the idea of $\Phi$-derivable approximations to include partially self-consistent schemes, and find sufficient criteria for the fulfillment of the continuity equation. We define a partially dressed functional $\Phi [G,G_0]$, in which the Green's function lines can be dressed with either $G$ or additional, fixed, Green's functions $G_0$. The fixed Green's functions could be outputs from a Hartree-Fock or density-functional theory calculation, but this is in no way necessary. In fact, sets of different fixed Green's functions can be used. We define the corresponding self-energy $\Sigma[G,G_0]$ via 
\begin{align}
 \Sigma [G,G_0] (1,2) = \left . \frac{\delta \Phi [G,G_0]}{\delta G(2,1^+)} \right |_{G_0 \text{ fixed}}. \label{self-energy2}
\end{align}
Solving the Kadanoff-Baym equations, \Eq{KBE1} and \Eq{KBE2}, with this self-energy defines a partially self-consistent scheme. We will refer to such schemes as partially $\Phi$-derivable. Dressing all Green's function lines in $\Phi$ with $G$ gives back the original $\Phi[G]$-functional as defined by Baym. Graphically, we obtain $\Sigma[G,G_0]$ from cutting away a $G$-line from all $\Phi[G,G_0]$-diagrams in every possible way. Examples of partially dressed $\Phi$-functionals and the corresponding self-energies are shown in \Fig{fig:ringDiagrams}, \Fig{fig:2ndOrderExchange} and \Fig{fig:fulfillingNumberConservation}.

In \Fig{fig:ringDiagrams} we show the example of a ring $\Phi$-diagram dressed in various ways. By dressing only one Green's function line with the full $G$, we obtain non-selfconsistent (one-shot, or single-shot) approximations. Combining \Fig{fig:ringDiagrams} a) and \Fig{fig:2ndOrderExchange}d), we obtain one-shot 2nd Born. Moreover, \Fig{fig:ringDiagrams} a) is part of the $G_0 W_0$ approximation. The $G_0 W_0$ approximation can be obtained by considering all higher order ring diagrams dressed with only one $G$. If all lines except one are dressed with $G$, as in \Fig{fig:ringDiagrams} c), we obtain several types of diagrams. In one diagram the baseline in $\Sigma$ is fully dressed, while in the other diagrams the baseline is not. Any one of the $\Sigma$-diagram in \Fig{fig:ringDiagrams} c) is by itself not partially $\Phi$-derivable, since closing it with a $G$-line and subsequently differentiating yields also the two other $\Sigma$-diagrams. The first of the $\Sigma$ diagrams in \Fig{fig:ringDiagrams} c) is part of the $G_0 W$ approximation. It is then readily seen that the $G_0W$ approximation for $\Sigma$ is not partially $\Phi$-derivable. The same can be seen in \Fig{fig:fulfillingNumberConservation}b), where we get two non-equivalent classes of $\Sigma$-diagrams from one $\Phi$-diagram. Each separate $\Sigma$-diagram is not partially $\Phi$-derivable, but their sum is. 

An important class of diagrams is obtained when each loop in $\Phi$ is dressed with either all $G$ or all $G_0$ separately. We will refer to this as a consistent dressing of the loops. Cutting  a $G$-line from consistently dressed $\Phi$-diagrams yields $\Sigma[G,G_0]$-diagrams with fully dressed baselines. An example of a consistently dressed $\Phi$-diagram is shown in \Fig{fig:ringDiagrams} b). This diagram is part of the $G W_0$ approximation. The remaining terms in the $G W_0$ approximation can be obtained by considering all higher order ring diagrams with only one loop consistently dressed with $G$, see also \Fig{fig:fulfillingNumberConservation}a). In the next section, we will show that all approximations coming from consistently dressed $\Phi$-functionals are number conserving. 


In a given $\Phi$-derivable approximation, the diagrammatic structure of $\Phi[G,G_0]$ is similar to the one for the fully dressed $\Phi[G]$
\begin{align}
 \Phi[G,G_0] = \sum _{nk} c_{nk} \int d1 d2 \Sigma_{k}^{(n)}[G,G_0] (1,2) G(2,1^+)
\end{align}
where $c_{nk}$ is a symmetry prefactor, $n$  is the number of interaction lines and $k$ labels the $\Sigma$ diagrams. For a collection of $\Sigma_k$ diagrams obtained from the same $\Phi$-diagram, the number $c_{nk}$ is generally given by one divided by the number of $G$-lines in $\Phi$. The proof of this statement is analogous to the proof that $c_{nk} = 1/2n$ in case that all $2n$ $G$-lines are dressed~\cite{stefanucci2013}. As an example, in FIG. \ref{fig:fulfillingNumberConservation}a) the $GW_0$ diagram has a prefactor of $1/2$, while the $T$-matrix diagram has $1/4$. In fact, for $GW_0$ each $\Phi$-diagram has the same prefactor $1/2$, independent on the order of the diagram. Note that the Feynman rules for $\Phi$ include a symmetry prefactor as well, related to the dimension of the symmetry group of each diagram~\cite{stefanucci2013}.

\subsection{Number conservation}
As was shown by Baym \cite{Baym1962} for fully dressed $\Phi$-functionals, particle number conservation is guaranteed by the inherent gauge invariance in $\Phi[G]$. Here, we repeat the same steps for partially dressed $\Phi[G,G_0]$-functionals.

Let us consider a change in $G$ of the form $G_\Lambda(1,2) = e^{i\Lambda(1)} G(1,2) e^{-i\Lambda(2)}$, where $\Lambda(t_{0}) = \Lambda(t_0 - i\beta)$ in order to satisfy the correct boundary conditions. For the exact Green's function, this transformation results from a gauge transformation $\Lambda (1)$~\cite{stefanucci2013}. The $\Phi$-functional will change according to $\Phi[G_\Lambda,G_0]$. For a general $\Phi$-functional, such as the ones in \Fig{fig:ringDiagrams} a) and c), $\Phi[G_\Lambda,G_0]$ will be different from $\Phi[G,G_0]$, and thus those $\Phi$ are not gauge invariant. However, a gauge invariant $\Phi$ is obtained if each loop is consistently dressed, such as in \Fig{fig:ringDiagrams} b) and d), since the phase factors always cancel for each loop separately. We thus have $\Phi[G_\Lambda,G_0] = \Phi[G,G_0]$, and hence $\delta \Phi = 0$ under gauge transformations.

For an arbitrary variation in $G(1,2)$, by definition the corresponding $\Phi$-functional will change according to 
\begin{align}
 \delta \Phi = \int d1 d2 \Sigma (1,2) \delta G (2,1^+).
 \label{deltaPhi}
\end{align}
For an infinitesimal gauge, 
\begin{align}
 \delta G(1,2) = G_\Lambda(1,2) - G(1,2) = i(\Lambda(1) - \Lambda(2) ) G(1,2).
\end{align}
the variation in $\Phi$ becomes
\begin{align}
 \delta \Phi = i\int d1 d2 \parenth{ \Sigma (1,2)G(2,1^+) - G(1,2)\Sigma (2,1^+)} \Lambda(1).
\end{align}
For gauge-invariant $\Phi$-functionals, $\delta \Phi = 0$ for any gauge $\Lambda$,  and thus we have
\begin{align}
 0 = \int d3 \commutator{ \Sigma (1,3)G(3,1^+) - G(1,3)\Sigma (3,1^+)}.
\end{align}
This is the source term which appeared in the continuity equation, \Eq{continuityEquation}. Thus, a gauge-invariant $\Phi$ leads to an approximation that conserves the particle number. If the approximate $\Phi$ used is not gauge invariant, the source term will in general be non-zero, leading to a violation of the particle number.

Since the resulting $\Sigma$ from a gauge-invariant $\Phi$ has a fully dressed baseline and consistently dressed loops, we can also infer that $\Sigma[G_\Lambda,G_0](1,2) = e^{i\Lambda(1)} \Sigma[G,G_0](1,2) e^{-i\Lambda(2)}$. $G_\Lambda$ and $\Sigma[G_\Lambda,G_0]$ fulfills the Kadanoff-Baym equations, \Eq{KBE1} and \Eq{KBE2}, with a Hamiltonian shifted with the gauge. One can then use the same considerations as in Ref.~\cite{stefanucci2013} to show that the Ward identity is satisfied, which then also leads to the fulfillment of the frequency sum rule for the density response function. As in the fully conserving case, this is however only valid under additional assumptions on the correct analytical structure of the response functions\cite{stefanucci2013,Hellgren2009}.

An approximate, gauge invariant, $\Phi$-functional is guaranteed to yield an approximation that fulfill the continuity equation for all times, independent on the particular shape of the time dependence in the Hamiltonian. For example, the relation is valid for a time-dependent single-particle potential, such as a time-dependent electric bias, relevant for quantum transport. The importance of using particle conserving approximations is exemplified by non-selfconsistent 2nd Born and $G_0W_0$, which can severely violate the continuity equation in quantum transport. \cite{Hershfield1992,Thygesen2008} 

For gauge-dependent approximations,  particle conservation can be violated already in ground-state calculations. A clear example can be found in Ref.~\cite{Stan2009a}, where the $H_2$ molecule was considered, starting with 2 particles in the non-interacting ground state. After an adiabatic switching, it was found that the particle number depended on the molecular bond length for $G_0W_0$, while it did not for $GW_0$. 

We also stress that any linear combination of consistently dressed $\Phi$-functionals generates particle number conserving approximations. One can then generate particle number conserving approximations when combining many-body methods, such as $GW$ + T-matrix~\cite{Springer1998} and the FLEX approximation~\cite{Bickers1989a,Bickers1989}, provided that each $\Phi$-diagram is consistently dressed.

Finally, we stress that gauge-invariant partial $\Phi$-derivability guarantees number conservation, but not other types of conservation laws ensured by a fully conserving scheme, such as momentum, angular momentum, and energy conservation\cite{Baym1962}. For example, the conservation of energy depends on the time-invariance of all four Green's function lines joining an interaction  line, and as such it is not enough for each loop to be dressed consistently separately. Furthermore, no mention has been done about how large the violations of the conservation laws are, since this also depend on how strongly correlated the system is~\cite{Schindlmayr1997}. In equilibrium, the violation of particle number seem to be small in some systems\cite{Schindlmayr2001,Stan2009a}, but larger for other systems,\cite{Schindlmayr1997} see also Sec.~\ref{AndersonExample}. In biased systems in quantum transport, the violation in the particle current can be as large as the current itself \cite{Hershfield1992, Thygesen2008}.

\section{Analytical properties of  \texorpdfstring{$G$}{G} and \texorpdfstring{$\Sigma$}{S}}\label{PSD}

\subsection{Analytical continuation and causality}

In the previous section we discussed particle number conserving approximations. In our discussion the particle number was defined by Eq.(\ref{N_particle})
using the Green's function in real time. In the following we will discuss only the equilibrium situation.
In that case there are at least two others ways to calculate the particle number. The first one is from an integral over the spectral function, as we will discuss in
much more detail below. Another way is from the derivative of the grand canonical potential $\Omega$ with respect to the chemical potential $\mu$ as
\begin{equation}
N = -\frac{\partial \Omega}{\partial \mu}.
\label{grandpotential}
\end{equation}
Baym showed that  for a conserving approximation the latter equation yields the same result as Eq.(\ref{N_particle}).
In his derivation he, however, implicitly assumed some analytic requirements of the Green's function, which are not guaranteed to be fulfilled.
In the reminder of the paper we will show that all three ways of calculating the particle number yield the same results for partially $\Phi$-derivable approximations provided that they preserve the correct complex analytic properties. We will further discuss a generalized sum rule for the particle number, the validity of which depends crucially on
the analyticity of the Green's function. 
For these reasons we will in this section first review the analytic properties of the Green's function and how they relate to the positive semi-definiteness of the spectral function.

For systems in equilibrium at inverse temperature $\beta$, the integrals over the $\gamma$ contour reduces to integrals over the imaginary axis. We can then make use of the Matsubara representation
\begin{align}
 \Gh^M (\tau_1,\tau_2) = \frac{1}{-i\beta} \sum _{m=-\infty}^\infty e^{-\omega_m (\tau_1 - \tau_2)} \Gh (\omega_m)
\end{align}
with Matsubara frequencies $\omega_m = \frac{ 2m+1}{-i\beta}$. The exact Green's function can be rewritten using the Lehmann representation\cite{Fetter2003} as 
 \begin{align}
 \Gh^M(\omega_n) = 
 \int _{-\infty}^\infty d\omega' \frac{\Ah (\omega')}{\omega_n + \mu - \omega'},
 \label{GMspectral}
\end{align}
where $\Ah(\omega)$  is the spectral function. 

We are interested in the analytical continuation of this function to the complex frequency plane. It is not obvious that such a continuation is unique, since $\Gh^M(\omega_m)$ is only defined on isolated points. Nevertheless, if we restrict the continuation to be analytic and bounded at infinity, the continuation is unique, and is given by \cite{Baym1961a}
\begin{align}
 \Gh^M (\zeta) &= 
 \int _{-\infty}^\infty d\omega' \frac{\Ah (\omega')}{\zeta + \mu - \omega'}, \quad \Im \zeta \neq 0. \label{GMSpectral}
 \end{align}
We use $\zeta$ as a complex frequency, while real frequencies are denoted by $\omega$. We note that \Eq{GMSpectral} immediately implies that $\Gh^M$ is analytic
away from the real axis.
\Eq{GMSpectral} can be used to define the commonly used retarded and advanced Green's functions in their respective half-planes as 
\begin{align}
 \Gh^M(\zeta) =
 \begin{cases}
 \Gh^R(\zeta+\mu) \quad \Im \zeta > 0   \\
 \Gh^A(\zeta+\mu) \quad \Im \zeta < 0.
 \end{cases} \label{GRetAnalytical}
\end{align}
The real-frequency retarded and advanced Green's functions are then obtained as $\Gh^M(\omega \pm i\eta) = \Gh^{R/A}(\omega + \mu)$, where $(\Gh^A)^\dagger (\omega) = \Gh^R (\omega)$ and $\eta$ a positive infinitesimal. $\Gh^M(\zeta)$ is discontinuous when crossing the real axis, with the discontinuity given by
\begin{equation}
\Gh^R(\omega) - \Gh^A(\omega) = -2\pi i \Ah(\omega),
\label{Aspectral}
\end{equation}
as can be deduced from Eq.(\ref{GMspectral}).
The functions $\Gh^{R/A}(\zeta)$ are analytic in their respective half-planes, which is what we will refer to as the correct analytical properties. 

The discussion so far has been independent of the self-energy. However, in practice, we often start from an equation of the self-energy. We therefore
want to relate the analytical properties of $G$ to those of $\Sigma$. This connection is provided by the Dyson equation, \Eq{Dyson}, 
which in terms of Matsubara notation becomes
\begin{align}
 \Gh^M(\omega_m) = \frac{1}{\omega_m + \mu - \hh - \Sigmah^M(\omega_m)},
\end{align}
where for operators $\hat{A}$ we use the notation $\hat{A}^{-1}=1/ \hat{A}$.
\cite{Baym1961a}

The same considerations for $\Gh^M(\zeta)$ also applies to the Matsubara self-energy.
Its analytic continuation is 
\begin{align}
 \Sigmah^M (\zeta) = 
 \int _{-\infty}^\infty d\omega' \frac{\hat{\Gamma} (\omega')}{\zeta + \mu - \omega'}, \quad \Im \zeta \neq 0, \label{SigmaSpectral}
\end{align}
where $\Gammah(\omega) $ is the rate operator. $\Sigmah^{R/A}(\zeta)$ are defined as in \Eq{GRetAnalytical}. Furthermore, $\Sigmah^M(\omega\pm i\eta) = \Sigmah^{R/A}(\omega+\mu)$, and $\Sigmah^{R}(\omega) - \Sigmah^{A}(\omega) = -2\pi i\Gammah(\omega)$. 
The analytic continuations $\Gh^M(\zeta)$ and $\Sigmah^M(\zeta)$ are related by
\begin{align}
  \Gh^M(\zeta) &= \frac{1}{\zeta + \mu - \hh - \Sigmah^M(\zeta)}, \quad \Im \zeta \neq 0. \label{GMDyson}
\end{align}

The discussion so far considered the exact Green's function and self-energy. However, in practice we typically approximate the 
self-energy and solve the Dyson equation. In a given approximation for $\Sigma$ the Green's function does not necessarily have the correct analytical
properties. For instance, the denominator in Eq.(\ref{GMDyson}) may have zeros away from the real axis in which case $\Gh^M(\zeta)$ has poles away from the
real axis and consequently a representation as in Eq.(\ref{GMspectral}) does not exist. However, such non-analyticities are not possible if the approximate self-energy is of the form
of Eq.(\ref{SigmaSpectral}) in which the rate operator $\hat{\Gamma} (\omega)$ is Positive Semi-Definite (PSD). With this we mean that $\langle \varphi | \hat{\Gamma} (\omega)| \varphi \rangle \geq 0$ for any one-particle basis function $\varphi$. In this case one can show that
$\Gh^M(\zeta)$ is analytic away from the real axis and has a representation as in Eq.(\ref{GMspectral}) with $\hat{A} (\omega)$ being PSD as well~\cite{Stefanucci2014}\cite{StefanucciErratum}\cite{Uimonen2015}. For this reason a diagrammatic perturbation method based on so-called half-diagrams has been devised in order
to guarantee that the rate function is always PSD~\cite{Stefanucci2014}\cite{StefanucciErratum}\cite{Uimonen2015}.

In case an approximate form for $\Sigmah^M$ does give rise to poles for $\Gh^M$ a generalized spectral representation, as in Eq.(\ref{GMspectral}), is possible.
In Appendix~\ref{AppendixC} we show that if $\Gh^M$ is analytic except for simple poles away from the real axis we can write
\begin{align}
\Gh^M (\zeta - \mu) =  \int_{-\infty}^\infty \! \! \! d\omega \frac{\hat{A}(\omega)}{\zeta-\omega}  + 
\sum_l \left (  \frac{\hat{\alpha}_l}{\zeta - \xi_l^R} \! + \!
  \frac{\hat{\alpha}^\dagger_l}{\zeta - (\xi_l^R)^*} \right )\! \!,\label{spectralandpoles}
\end{align}
where $\hat{\alpha}_l$ is the residue matrix, $\xi^R_l$ is the location of a pole in the upper half-plane, and the spectral function $\Ah(\omega)$ is also  in this  case given by Eq.(\ref{Aspectral}). If $\Gh^M$ is obtained from Eq.(\ref{GMDyson}) then from taking the limit $|\zeta| \to \infty$ (assuming that the approximate
$\Sigma$ is bounded at infinity) we see from
\begin{align}
\Gh^M (\zeta) \to \frac{\oneh}{\zeta}  \quad  (|\zeta| \to \infty)
\end{align}
 that
\begin{align}
\oneh = \int_{-\infty}^\infty d\omega \hat{A} (\omega) + \sum_l \left ( \hat{\alpha}_l + \hat{\alpha}_l^\dagger \right),
\label{normChange}
\end{align}
where $\oneh$ is the unit operator. Thus, the presence of poles can change the norm of the spectral function.

\begin{figure*}
  \centering
   \includegraphics[width=0.48\textwidth]{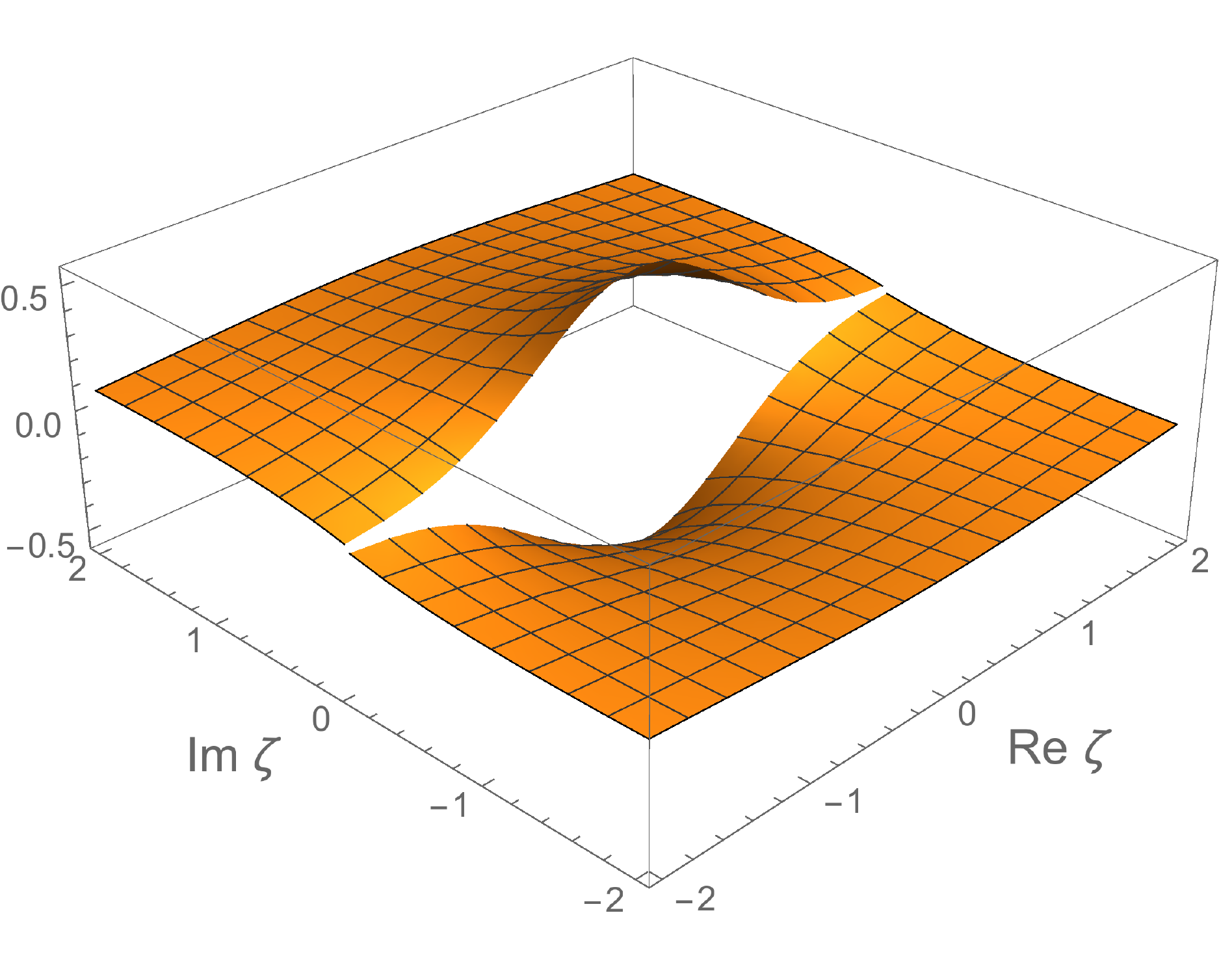}
   \includegraphics[width=0.48\textwidth]{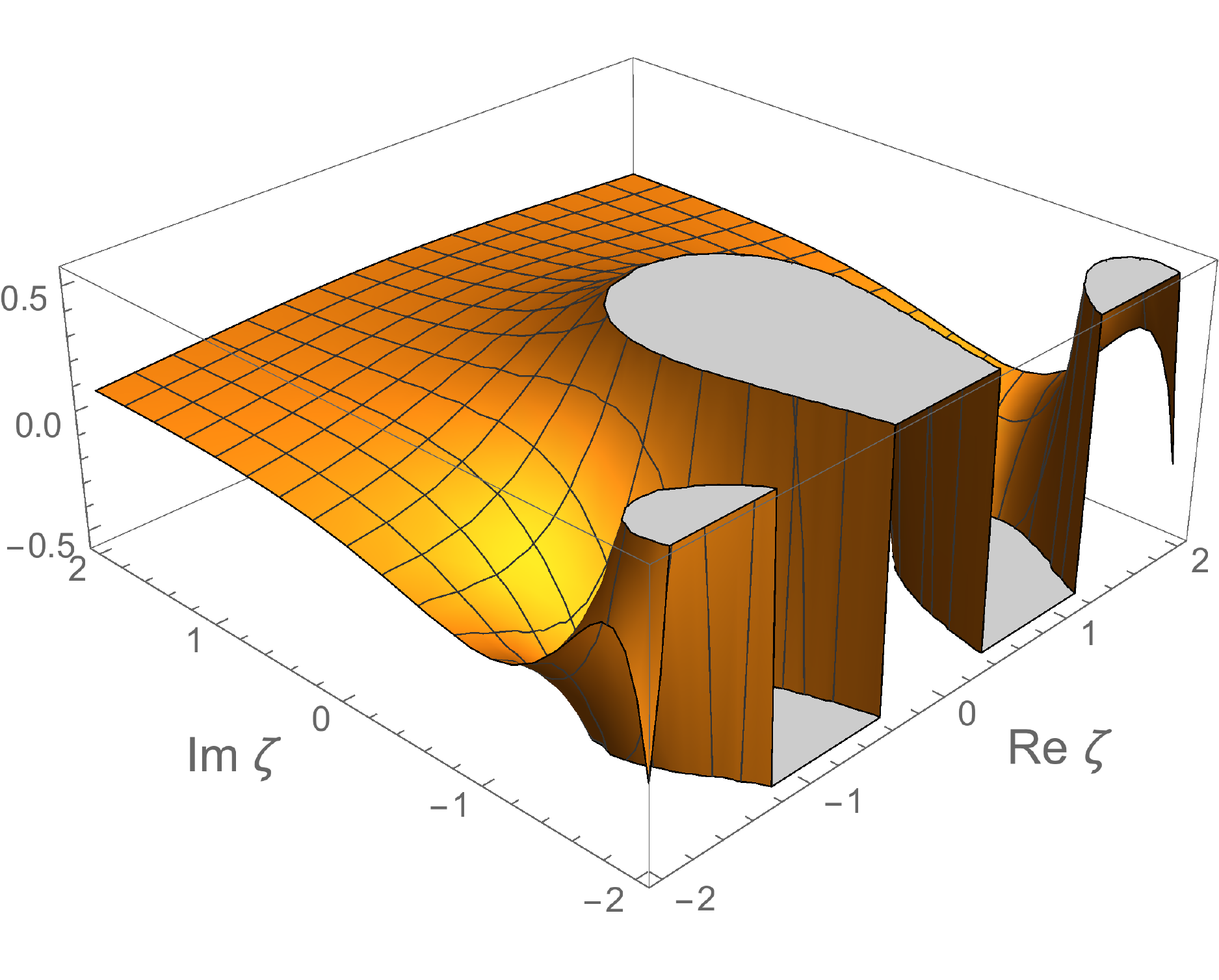}   
  \caption{Left: $-\frac{1}{\pi} \Im G^M(\zeta)$ for a Gaussian spectral function, \Eq{GRetGaussian}, with $a=1$. $G^M(\zeta) \to 1/\zeta$ for large $\zeta$. Right: Analytically continued $-\frac{1}{\pi} \Im G^R(\zeta)$ for a Gaussian spectral function, \Eq{GretContinued}, with $a=1$. $G^R(\zeta)$ is analytic everywhere and is unbounded in the lower half-plane. High and low values has been cut off for the sake of clarity. 
   \label{fig:gaussian} }
 \end{figure*}
While discussing the analytic properties, we would like to clear up a possible point of confusion regarding the analytic continuation of the
retarded and advanced functions $G^{R/A} (\zeta)$ from the upper/lower half planes in which they are analytic, across the real axis to the other 
half planes. Such continuations are useful for derivations with the help of Cauchy's theorem and have also been useful
for interpreting quasi-particle life-times in terms of analytic properties of the analytically continued $G^{R/A} (\zeta)$.

Due to the discontinuity in $G^M(\zeta)$, the analytical continuation from the upper half-plane, where $G^M(\zeta) = G^R(\zeta)$, into the lower half-plane yields a completely different function than $G^A(\zeta)$. 
The analytically continued $G^R(\zeta)$ is used in several textbooks (see, for example, Refs.~\cite{Fetter2003, Gross1991}) when the quasi-particle peak of the spectral function has a Lorentzian shape. A typical example is the homogeneous electron gas in which case we can regard the spectral function as a scalar function rather than a matrix, as it is diagonal in momentum space. We consider a Lorentzian with width $\Gamma$ as
\begin{equation}
A(\omega) = \frac{\Gamma}{\pi} \frac{1}{\omega^2 + \Gamma^2}.
\end{equation}
We obtain $G^{R} (\zeta) = \frac{1}{\zeta+ i \Gamma}, \Im \zeta > 0$ from \Eq{GMSpectral}. The analytical continuation is given by the same expression, but defined for all $\zeta$ and has a pole in the lower half-plane.  A useful application of the analytic continuation is
the calculation of the retarded Green's function in time-space, 
\begin{align}
 G^R(t) =\int _{-\infty} ^\infty \frac{d\omega}{2\pi} G^R(\omega) e^{-i\omega t}
\end{align}
by the following procedure. For $t<0$, $G^R(\zeta) e^{-i \zeta t}$ decays exponentially in the upper half-plane, and we can integrate around a closed half-circle in the upper half-plane. Since $G^R(\zeta)$ has no pole there, $G^R(t) =0$ for $t<0$, befitting of a retarded function. This also shows that the existence of poles in the upper half-plane will in general violate the causality property of $G^R(t)$.  For $t>0$, we integrate around a half-circle in the lower half-plane and find
\begin{align}
G^R(t) = -i \theta(t) e^{-\Gamma t}.
\end{align}
Thus, the location of the pole in the lower half-plane yields the lifetime of an added particle. This interpretation was entirely based on the Lorentzian form
of the quasi-particle peak of the spectral function. However, it is known for the homogeneous electron gas that the moment $\int d\omega \, \omega^2 A(\omega)$ of the spectral function
is finite, which precludes a Lorentzian form~\cite{Pavlyukh2013}. A study of the short-time properties of the spectral function~\cite{Bonitz1999,Pavlyukh2013} shows that for this system a more realistic form
of the retarded Green's function in real time is:
\begin{equation}
G^R (t) = -i \theta (t) \exp{ \left(- \gamma \frac{t^2}{t+ \tau} \right) } 
\end{equation}
with $\gamma$ and $\tau$ parameters. This Green's function has a Gaussian short-time behavior and an exponential long-time behavior. We would like to demonstrate that such a short-time
behavior can be described by a simple spectral function of Gaussian form, $A(\omega) = \frac{1}{a\sqrt{\pi}} e^{-\frac{\omega^2}{a^2}}$, where $a>0$ determines the width of the single peak, and concentrate on the analytical properties. From \Eq{GMSpectral}, we obtain 
\begin{align}
 G^M(\zeta) = e^{-\frac{\zeta^2}{a^2}}\frac{ \pi  \text{Erfi}(\zeta/a) + \ln \left(-\zeta \right)-\ln (\zeta) }
 {a\sqrt{\pi}}, \Im \zeta \neq 0, \label{GRetGaussian}
\end{align}
where $\text{Erfi}$ is the imaginary error function, analytic in the whole complex plane. As it should,  $G^M(\zeta) \to 1/\zeta$  as $|\zeta| \to \infty$. The analytically continued $G^R(\zeta)$ is given by 
\begin{align}
  G^R(\zeta) = \frac{\sqrt{\pi}}{a} e^{-\frac{\zeta^2}{a^2}} \left ( \text{Erfi}(\zeta/a) - i \right ),
  \label{GretContinued}
\end{align}
which is analytic for all $\zeta$ in the complex plane. Thus, for a Gaussian spectral function, $G^R(\zeta)$ does not have any poles in the lower half-plane, as opposed to the case with a Lorentzian spectral function. Moreover, $G^R(\zeta)$ is unbounded in the lower half-plane, where $G^R(\zeta)\to -\frac{2 i \sqrt{\pi } }{a}e^{-\frac{\zeta^2}{a^2}} $ for $|\zeta| \to \infty$.
This affects the calculation for $G^R(t)$ using contour integration. We can close a contour in the upper half-plane for $t<0$  yielding $G^R(t) = 0$, but we cannot close the contour in the lower half-plane for $t>0$, since the integral on the half-circle diverges, see FIG.~\ref{fig:gaussian}. Nevertheless, $G^R(t)$ is well-defined for all $t$, and can be obtained by Fourier transforming $G^R(\omega)$ directly, yielding
\begin{align}
 G^R(t) = -i \theta(t) e^{-\frac{1}{4} a^2 t^2}.
\end{align}
We see that, similarly to the Lorentzian case, the width of the spectral function determines the lifetime of an added particle excitation. However, in this case the lifetime is unrelated to the properties of $G^R(\zeta)$ analytically continued to the lower half-plane. The main message that we want to give in this section is therefore that the analytic continuations of the retarded and advanced functions beyond their original analytic domains in general can give rise to a richer analytical structure than just simple poles. This not only includes unbounded analytic functions, but, for instance, also functions with algebraic or logarithmic branch cuts.

\subsection{PSD and analyticity}\label{simpleModel}

One way of ensuring the correct analytical properties of $G^M(\zeta)$ is to ensure that the chosen diagrammatic approximation retain the PSD property. If the diagrammatic structure has this property, then the correct analytical properties are guaranteed \cite{Uimonen2015a}. Recently, it has been shown~\cite{Stefanucci2014,Uimonen2015} that by choosing diagrams in a specific way, by only keeping diagrams that can be built up from so-called half-diagrams, the PSD property and thus the correct analytical properties are guaranteed at zero temperature. Moreover, the PSD properties depend only on the diagrammatic structure, but not on the dressing of the loops provided that we dress with a $G$ coming from a PSD approximation. It is thus important to note that conservingness and PSD are two completely different properties, a point which we feel is not stressed enough in the literature. For example, $G_0W_0$ is not conserving, but PSD, while the fully dressed second-order exchange approximation (see FIG.\ref{fig:2ndOrderExchange}b)) is conserving, but not PSD.~\cite{Stefanucci2014} Thus, we are guaranteed to have the correct analytical properties for the former approximation, but not the latter.

Non-PSD spectral functions has been observed~\cite{Minnhagen1974,Stefanucci2014} in studies of the homogeneous electron gas, when one goes beyond $GW$ and takes vertex corrections into account. The lowest order vertex correction is the second-order exchange diagram, where the interaction lines have been dressed with the screened interaction.  Moreover, it has also been observed~\cite{Minnhagen1974} that in this approximation a pole can be created in the lower half-plane for $G^A(\zeta)$, violating the analytical properties. Also in finite systems, restricting to certain classes of diagrams can violate positivity for the photo-absorption spectra.~\cite{Hellgren2009,Sangalli2011}

To elucidate the above considerations, and to illustrate \Eq{spectralandpoles}, we consider a very simple model of an interacting self-energy. We assume that the analytically continued $\Sigma^R (\zeta)$ has a single pole on the real axis, with residue $\alpha \in \mathbb{R}$,
\begin{align}
 \Sigma^R (\zeta) = \frac{\alpha}{\zeta - b }.
\end{align}
$\Sigma^R(\zeta)$ is analytic in the upper half-plane, and $\alpha > 0 (<0)$ corresponds to a positive (negative) definite rate operator $\Gamma(\omega) = -1/\pi \Im [ \Sigma^R(\omega)]$ on the real axis. The resulting $G^R(\zeta)$ is 
\begin{align}
 G^R (\zeta) = \frac{1}{\zeta - \Sigma^R(\zeta) }.
\end{align}
which has two poles, at 
\begin{align}
  \omega_{\pm} = \frac{1}{2} \left(b \pm \sqrt{4 \alpha +b^2}\right).
\end{align}
If $4\alpha > -b^2$, $G^R(\zeta)$ is analytic in the upper and lower half-planes, but otherwise we obtain two poles in the separate half-planes. The imaginary part is 
\begin{align}
 \Im [\omega _{\pm}] = \begin{cases}
                        0, \quad &4\alpha > -b^2 \\
\pm \sqrt{-(4 \alpha +b^2)}, \quad &4\alpha < -b^2 
                       \end{cases}
\end{align}
\begin{figure}
 \centering
 \includegraphics[width=0.48\textwidth]{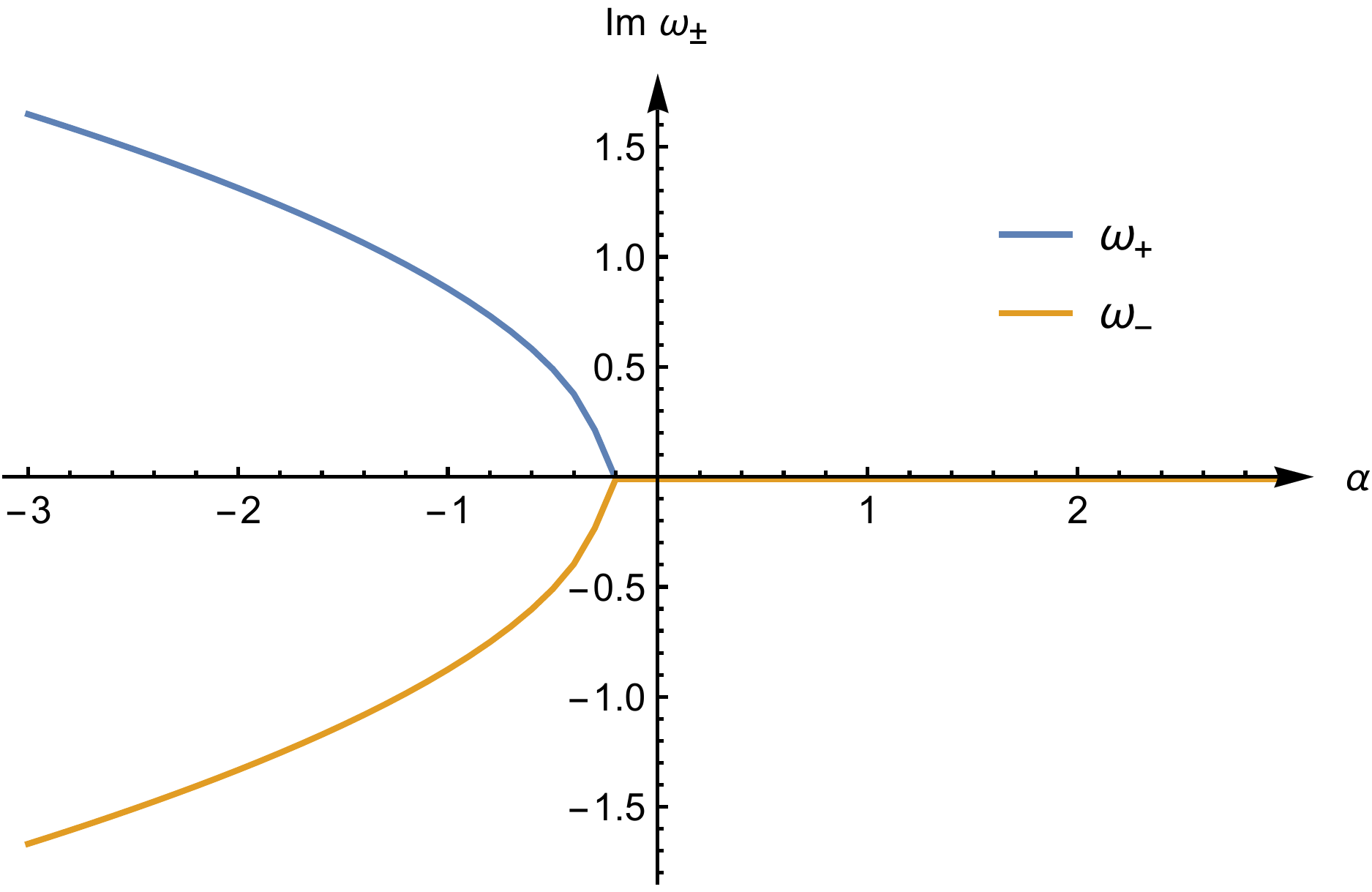}
  \includegraphics[width=0.48\textwidth]{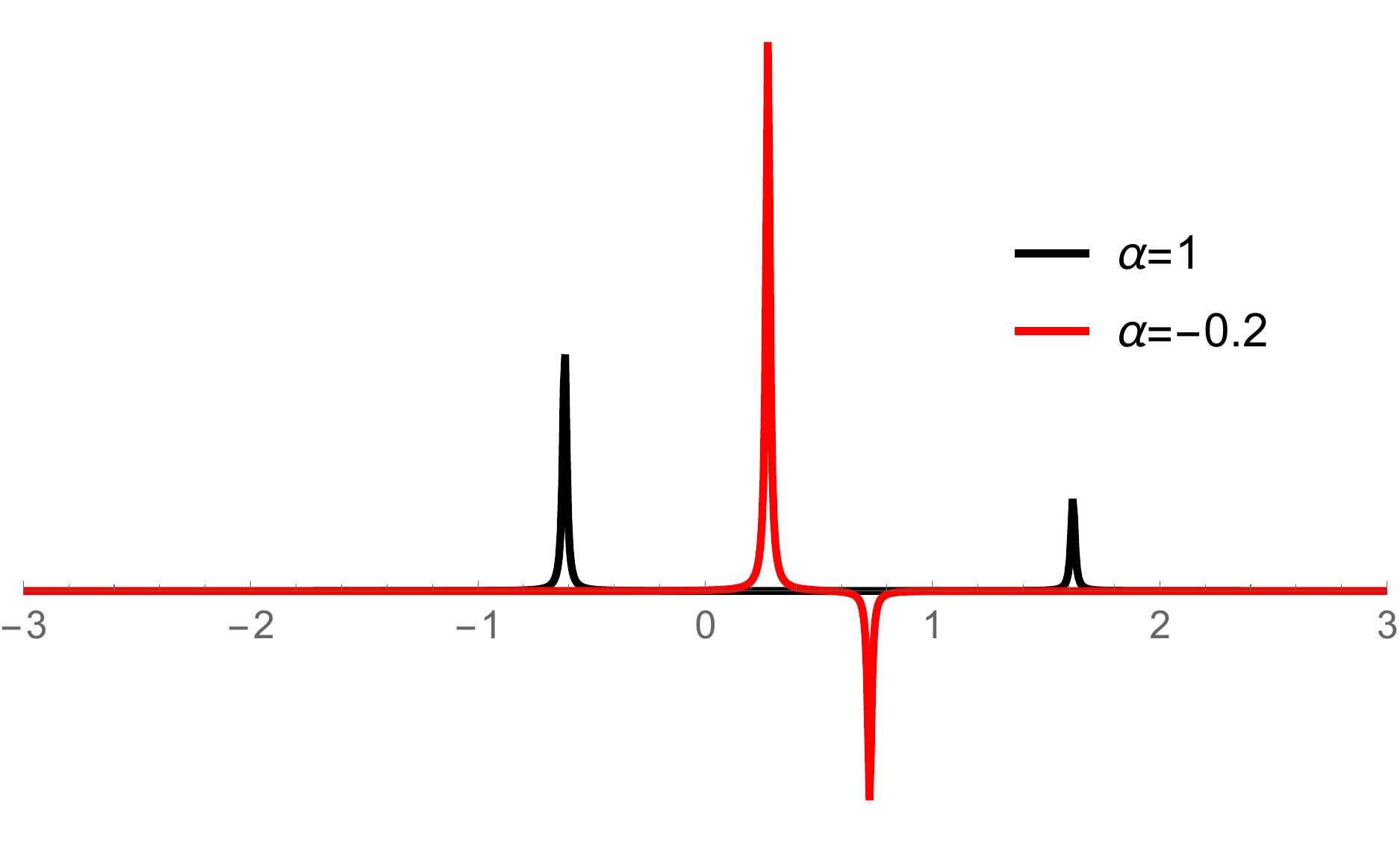}
 \caption{(Upper) The imaginary part of the two poles $\omega_{\pm}$ of $G^R$ as function of $\alpha$. The branching occurs at $4\alpha=-b^2$. (Lower) The spectral function $A(\omega)$ for $\alpha>0$, where $A(\omega) > 0$, and $-4b^2 <\alpha < 0$, where $A(\omega)$ can become negative. For $\alpha < -4b^2$, $A(\omega) = 0$, the delta function weights are zero.
 The parameters are $b=1$, with a finite broadening. }
 \label{fig:simpleModel}
\end{figure}
If $G^R(\zeta)$ is non-analytic in the upper half-plane, it is impossible to write $G^R(\zeta)$ in the form of \Eq{GMSpectral}. Nevertheless, we can define the spectral function as $A(\omega) =-\frac{1}{\pi} \Im G^R(\omega)$ due to \Eq{spectralandpoles}, which will be the sum of two delta-functions in case the poles are on the real axis. Now, however, the integral over the spectral function will depend on the position on the pole. The integral can be obtained by integrating $G^R(\zeta)$ over a semi-circle in the upper half-plane. In this model, $G^R(\zeta) \to 1/\zeta, \zeta \to \infty$, and thus the half-circle integration yields $\pi i$. If $4\alpha < -b^2$, then there is a pole $\omega_+$ in the upper half-plane, with residue
\begin{align}
 \alpha_1= \frac{1}{2} - i\frac{b}{2 \sqrt{-(4 \alpha +b^2)}}
\end{align}
This gives the integral as 
\begin{align}
 \int_{-\infty} ^\infty A(\omega) d\omega = 
 \begin{cases}
 1 \quad 4\alpha > -b^2 \\
 0 \quad 4\alpha < -b^2 \\
 \end{cases}
\end{align}

The spectral function is plotted in \Fig{fig:simpleModel} for different values of $\alpha$. As can be seen, if $\alpha > 0$, $A(\omega)$ is positive definite. If $-b^2 <4\alpha < 0$, $A(\omega)$ has one negative and one positive peak. In this simple model, we can thus identify three regions. 
\begin{itemize}
 \item $\alpha > 0$. The rate operator is positive definite, and thus the spectral function is positive definite. This leads to that $G^R(\zeta)$ is analytic in the upper half-plane,  and $\int d\omega A(\omega) = 1$.
 \item $-b^2 <4\alpha < 0$. The rate operator is negative definite, and $A(\omega)$ can have both positive and negative weights. $G^R(\zeta)$ is analytic in the upper half-plane, and $\int d\omega A(\omega) = 1$. 
 \item $4\alpha < -b^2$. The rate operator is negative definite, and $A(\omega) = 0$. There is a pole in the upper half-plane for $G^R(\zeta)$ and is of the form of \Eq{spectralandpoles} in the upper half-plane;
 \begin{align}
  G^R(\zeta) = \frac{\alpha_1}{\zeta - \xi^R_1} + \frac{\alpha_1^*}{\zeta - (\xi^R_1)^*},
 \end{align}
  where $\xi^R_1 = \omega_+$ and $(\xi^R_1)^* = \omega_-$. 
\end{itemize}

Keeping in mind that $\Sigma^R(\zeta)$ is a very simple model of an interacting self-energy, the case $\alpha > 0$ corresponds to when we can write the self-energy as a sum of squares of half-diagrams\cite{Uimonen2015,Uimonen2015a}. The case $\alpha < 0$ corresponds to when the self-energy is not a sum of squares of half-diagrams. However, we see that it is not guaranteed that the correct analytic properties are broken. Taking $\alpha$ as a measure of the correlation effects, for $\alpha < 0$ we find a critical point when the analytical properties are broken. While being a simple model, the same trend can be observed in the Anderson model, discussed in Sec.~\ref{AndersonExample}.

\section{Gauge invariance and sum rules} \label{sumRules}
Gauge invariant $\Phi$-functionals leads to approximations that obey well-known sum rules in equilibrium. Such sum rules have been discussed before (see, e.g. \cite{Luttinger1960a, Shiba1975, Yoshimori1982}). Here, we repeat the derivations for approximations that are partially $\Phi$-derivable, but we keep a general structure and focus on the importance of the analytical  properties.

We consider the number of particles in equilibrium, given by the Matsubara Green's function as 
\begin{align}
 N &= - i \trace{\Gh^M (\tau, \tau^+)} 
 = \frac{1}{\beta} \trace{ \sum _{m=-\infty} ^\infty e^{\eta \omega_m} \Gh^M(\omega_n) }= \nonumber
 \\
 &= \frac{1}{\beta}  \trace{\sum _{m=-\infty} ^\infty e^{\eta \omega_m} \frac{-1}{
 \Sigmah ^M (\omega_m) + \hh - \mu - \omega_m } }, \label{matsubara}
\end{align}
where $\eta \to 0^+$ after the summation. According to our previous discussion, we can analytically continue $\Gh^M(\omega_m)$ to $\Gh^M(\zeta)$ when $\Im \zeta \neq 0$, for any finite temperature. Here, we wish to point out an issue regarding the analytical properties. 

Matsubara sums can be rewritten according to the following rule (see, for example, Ref.\cite{Fetter2003}), which follows from considering two half-circles, in the upper and lower half-plane, when $\hat{Q}(\zeta)$ is analytic for $\Im \zeta \neq 0$,
\begin{align}
\begin{split}
 \frac{i}{\beta} \sum _{m=-\infty} ^\infty e^{\eta \omega_m} \hat{Q} (\omega_m) = \\
  = \lim _{\eta \to 0^+} \int _{-\infty} ^{\infty} \frac{d\omega}{2\pi} f(\omega) \left ( \hat{Q} (\omega - i\eta) - \hat{Q} (\omega + i\eta)\right ),
\end{split}
  \label{eq:needAnalytic}
\end{align}
where $f(\omega) = 1/(e^{\beta \omega  } +1)$ is the Fermi function.
By taking $\hat{Q}=\Gh^M$ and using Eq.(\ref{matsubara}), we can write $N$ as
\begin{align}
\begin{split}
 N &= \int _{-\infty} ^{\infty} \frac{d\omega}{2\pi i} f(\omega) \trace{ \Gh^A (\omega + \mu) - \Gh^R (\omega +\mu) }= \\
&= \int _{-\infty} ^{\infty} d\omega f(\omega - \mu) \trace{\Ah(\omega)} \label{particleNumber2} 
\end{split}
\end{align}
\emph{provided} that $\Gh^M(\zeta)$ is analytic when $\zeta \neq 0$. If $\Gh^M(\zeta)$ would have simple poles, we must add their residues to \Eq{particleNumber2}
as these poles give an extra contribution in the derivation of Eq.{\ref{eq:needAnalytic}}.
We  conclude, that if $\Gh^M(\zeta)$ is non-analytic, then  $- i \trace{\Gh^M (\tau, \tau^+)} $ and $\int _{-\infty} ^{\infty} d\omega f(\omega - \mu) \trace{\Ah(\omega)}$ can give different results. This shows yet another inconsistency if one works with approximations that do not guarantee the correct analytical properties. Note, however, that no assumption was made on the existence of a generating $\Phi$-functional. 

Yet another way of obtaining the particle number is via the grand canonical potential, as was considered by Baym~\cite{Baym1962} for conserving schemes, via \Eq{grandpotential}. For a partially $\Phi$-derivable approximation with the correct analytical properties, this give the same particle number as the two other definitions discussed. For details, see Appendix~\ref{AppendixD}.

We now focus on rewriting \Eq{matsubara}. Since the continuation of $\Gh^M(\omega_m)$ to $\Gh^M(\zeta)$ is analytic at the Matsubara frequencies, we can perform derivatives around $\omega_m$. We can thus write \Eq{matsubara} as 
\begin{widetext}
\begin{align}
 N =   
   \underbrace{ \frac{1}{\beta} \sum _{m=-\infty} ^\infty e^{\eta \omega_m} 
   \trace{ \left . \frac{\partial }{\partial \zeta} \ln \left [ \Sigmah ^M (\zeta) + \hh - \mu - \zeta \right ] \right |_{\zeta=\omega_m}}}_{I_1}
   + 
   \underbrace{\frac{1}{\beta} \sum _{m=-\infty} ^\infty e^{\eta \omega_m} 
   \trace{\Gh^M (\omega_m) \left . \frac{\partial \Sigmah^M}{\partial \zeta}\right |_{\zeta=\omega_m} }}_{I_2}.
   \label{particleNumber}
\end{align}
\end{widetext}
where we have defined $I_1$ and $I_2$, and taken the branch cut of the complex logarithm to be the negative real axis. The relation used for the matrix logarithm,
\begin{align}
 \frac{\partial}{\partial \omega} \trace{ \ln \left [ \Qh(\omega) \right ] } = \trace{\Qh^{-1}(\omega) \frac{\partial}{\partial \omega} \Qh(\omega)},
 \label{derivativeMatrix}
\end{align}
is valid for all $\Qh(\omega)$ that are smooth and invertible but not necessarily diagonalizable, see Appendix~\ref{AppendixB}. The reason for introducing $I_1$ and $I_2$, is that at zero temperature $I_2$ is related to variations of the $\Phi$-functional, and is zero for number-conserving approximations. Furthermore, $I_1$ can be integrated analytically.

\Eq{particleNumber} simplifies at zero temperature. We will first discuss $I_2$. For small temperature $T$, the distance between the Matsubara frequencies, $d\omega = 2\pi i / \beta$, becomes small. In the limit $T\to 0$, a Matsubara sum can be written as an integral over the upper and lower half-plane,
\begin{align}
 \lim_{T\to 0} \frac{1}{\beta} \sum _{m=-\infty} ^\infty e^{\eta \omega_m} \hat{Q} (\omega_m) = 
  P\int _{-i\infty} ^{i\infty} \frac{d\zeta}{2\pi i} \hat{Q} (\zeta) ,
 \label{zeroTemp}
\end{align}
where the principal value of the integral is \mbox{$P\int_{-i\infty} ^{i\infty} = \lim_{\epsilon \to 0^+} \int_{-i\infty} ^{-i\epsilon} + \int_{i\epsilon} ^{i\infty}$}. \Eq{zeroTemp} is valid if $\Qh(\omega_m)$ is not too singular around $0$, see Appendix~\ref{AppendixA}, and also the discussion in Ref.~\cite{Luttinger1960}. With this relation, we can write $I_2$ as $T\to 0$ as 
\begin{align}
  I_2 = \int _{-i\infty} ^{i\infty} \frac{d\zeta}{2\pi i} \trace{\Gh^M (\zeta) \frac{\partial \Sigmah^M}{\partial \zeta} }. \label{I2identity}
\end{align}

In \Eq{I2identity} we have made the assumption of integrability and replaced the principal value of the integral by the  integral itself. We will here discuss these assumptions. $\Gh^M(\zeta)$ is analytic and thus integrable  for $\Im \zeta \neq 0$, and has a discontinuity around $\zeta = 0$ given by the spectral function at $\mu$, \Eq{Aspectral}. Thus, assuming that $\Ah (\mu)$ is finite, $\Gh^M(\zeta)$ is integrable. By the same argument, $\Sigmah^M(\zeta)$ and $\frac{\partial \Sigmah^M}{\partial \zeta}$ are integrable for $\Im \zeta \neq 0$. 
From \Eq{SigmaSpectral}, 
\begin{align}
 \left . \frac{\partial \Sigmah}{\partial \zeta}\right |_{\zeta=0} = -\int _{-\infty}^\infty d\omega' \frac{\hat{\Gamma} (\omega')}{ (\mu - \omega')^2},\label{derivSigma}
\end{align}
from which we see that if $\Gammah(\omega)$ decays quickly enough at $\mu$, $\frac{\partial \Sigmah}{\partial \zeta}$ is continuous also at $\zeta = 0$. For Fermi liquids, $\Gammah(\mu) \sim (\omega-\mu)^2$, for $\omega$ close to $\mu$,~\cite{Luttinger1961} and as such \Eq{derivSigma} is integrable. However, for other systems, such as the one-dimensional electron gas~\cite{Giuliani2005}, or in Mott insulators~\cite{Rusakov2016}, the rate operator can decay slower or even diverge at $\mu$. Excluding such cases, the integral in \Eq{I2identity} is well-defined, and equal to its principal value. 

$I_2$, also referred to as the Luttinger integral, can be related to the $\Phi$-functional under special types of variations in $G$, namely those that shift the Matsubara frequencies. 
As an example, consider the consistently dressed ring diagram in FIG \ref{fig:ringDiagrams} b). $\Phi[G,G_0]$ is then given as
\begin{align}
\begin{split}
 \Phi = \! -\frac{i^2}{2} \! \! \int d1d2 G(1,2) G(2,1) v(1,2)^2 G_0(1,2) G_0(2,1) = \\
      = \! -\frac{1}{2\beta^2} \! \! \sum _{nmrq} \! \! \! G^M(\omega_n) G^M(\omega_m) v^2 G_0^M(\omega_r) G_0^M(\omega_q) \delta_{0,n-m+r-q}, \label{structurePhi}
\end{split}
\end{align}
where spatial integration is suppressed in the second line, to focus on the $\omega$-dependence. 
The Kronecker delta represents frequency conservation at every vertex, due to the time-locality of the interaction. $\Phi$ evaluated at a $G$ which has shifted frequencies, $G^M(\omega_m+\delta \omega)$, (defined via its analytical continuation, \Eq{GMSpectral}, as $G^M(\zeta)|_{\zeta = \omega_m+\delta \omega}$), would in general be different from the unshifted case. For the special case of a shift corresponding to one (or more) Matsubara frequency, i.e. $\delta \omega = \frac{2\pi i}{\beta}$ which gives $\omega_m + \delta \omega = \omega_{m+1}$, we can relabel the sum for $m$ and $n$ in \Eq{structurePhi}. Since we perform the same relabeling for both $m$ and $n$, the Kronecker delta does not change, which leaves $\Phi$ invariant. If the loop was not consistently dressed, $\Phi$ would change since the Kronecker delta would change. The argument works for a general diagram, and thus we see that if the loops are consistently dressed, $\delta \Phi = 0$ when $\omega_m \to \omega_{m+1}$ in $G^M$. Otherwise, $\delta \Phi \neq 0$ in general.

The reason why consistently dressed $\Phi$-diagrams are invariant under frequency shifts is due to the interaction being time-local. The loops in $\Phi$ can then be considered separately. This type of argument can not,  however, be used for other types of conservation laws. For example, momentum conservation comes from the invariance of $\Phi$ under spatial translations in $G$~\cite{Baym1962}, for interactions that depend only on the interparticle distance. For such a shift in $G$ in \Eq{structurePhi}, by a variable transform we only shift one index in the interaction, and as such $\Phi$ is not invariant. For a $\Phi$-diagram to be conserving, we need, in general, full $\Phi$-derivability.

Consider again the change in $\Phi$ when we make frequency shift in $G$. This is given by \Eq{deltaPhi}, which can be written using Matsubara Green's functions as 
\begin{align}
 \delta \Phi = \sum _{m=-\infty} ^\infty e^{\eta \omega_m} \trace{\Sigmah^M (\omega_m) \delta \Gh^M (\omega_m) }.
 \label{variationPhi}
\end{align}
Let us fix the shift to $\delta \omega = \frac{2 \pi i}{\beta}$, and take the zero-temperature limit of \Eq{variationPhi}. Here, we cannot use \Eq{zeroTemp} directly, since the integrand is too singular around 0 (see Appendix~\ref{AppendixA}). The limit is
\begin{align}
 \delta \Phi \underset{T \to 0}{\to} P \int _{-i\infty} ^{i\infty} \trace{\Sigmah^M (\omega) \frac{\partial \Gh^M}{\partial \omega} } d\omega + \\
  + \lim _{\epsilon \to 0^+} \trace{\Sigmah^M(0) \left [ \Gh^M(i\epsilon) - \Gh^M(-i\epsilon) \right ] }.
  \label{beforeWard}
\end{align}
By making use of partial integration, we obtain  
\begin{align}
 \delta \Phi \underset{T \to 0}{\to} -\int _{-i\infty} ^{i\infty} d\zeta \trace{\Gh^M (\zeta) \frac{\partial \Sigmah^M}{\partial \zeta} } = -2 \pi i I_2. \label{WardIdentity}
\end{align}
Thus, at zero temperature, we can relate $I_2$ to special types of variations of $\Phi$. For a partially $\Phi$-derivable approximation, $I_2 = 0$ if $\Phi$ is gauge invariant. Otherwise, $I_2 \neq 0$ in general. 

We now derive expressions for $I_1$ and $I_2$ involving integrals over real frequencies at zero temperature. Using \Eq{eq:needAnalytic} we obtain $I_2$ as
\begin{align}
   I_2 = -2\Im  \left [ \int _{-\infty} ^{\mu} \frac{d\omega}{2\pi} \trace{ 
   \Gh^R (\omega ) \frac{\partial \Sigmah^R}{\partial \omega} } \right ]. \label{I2ret}
\end{align}
\Eq{I2ret} can also be derived from \Eq{WardIdentity} by considering the contour integral around two quarter-circles in the upper and lower half-plane.

Using \Eq{eq:needAnalytic} for $I_1$, partial integration, and $\Sigmah^M(\pm i \eta) = \Sigmah^{R/A} (\mu)$, we obtain
\begin{widetext}
\begin{align}
 N =
   \underbrace{\frac{1}{2\pi i}
   \trace{ \ln \parenth{ \Sigmah^A(\mu) + \hh - \mu + i\eta } - \ln \parenth{ \Sigmah^R(\mu) + \hh - \mu - i\eta } }}_{I_1}
   \underbrace{-2\Im  \left [ \int _{-\infty} ^{\mu} \frac{d\omega}{2\pi} \trace{ 
   \Gh^R (\omega ) \frac{\partial \Sigmah^R}{\partial \omega} } \right ]}_{I_2}.
   \label{generalizedLuttinger}
\end{align}
\end{widetext}
The boundary terms for $I_1$ at $\omega \to -\infty$ gives $\Im [\ln (\infty \pm i\eta)] = 0$, which explains our choice of branch and the factor of $(-1)$ in \Eq{matsubara}.

In deriving \Eq{generalizedLuttinger} we have used the relation
\begin{align}
 \int _\infty ^\mu \! \! \frac{\partial}{\partial \omega} \trace{ \ln \left [ \Qh (\omega) \right ]} \! = \! 
 \trace{\ln \left [ \Qh(\mu) \right ] \! - \! \ln \left [\Qh(-\infty) \right ]}.
 \label{hopeNoBranchCut}
\end{align}
Here one has to assume that during the integration path, none of the eigenvalues $\lambda_k(\omega)$ of $\Qh(\omega)$ cross the branch cut of the logarithm, where we regard $\lambda_k(\omega)$ as a curve in the complex plane parametrized by $\omega$. This can be seen via 
\begin{align}
 \trace{\ln \left [ \Qh(\omega) \right] } = \sum_k \ln \left [ \lambda_k(\omega) \right ],
 \label{traceLog}
\end{align}
valid for any complex invertible matrix, see Appendix~\ref{AppendixB}. If one or more of the eigenvalues traverse the branch cut at $(-\infty,0)$ as function of $\omega$, a unique logarithm cannot be found, and the argument principle (see, e.g., Ref.~\cite{Gamelin2003}) has to be used instead of \Eq{hopeNoBranchCut}. Crossing the branch cut changes the logarithm with $2\pi i$ if it is crossed from above, and $-2\pi i$ if it is crossed from below. 
However, if the rate operator is PSD no eigenvalue can cross the real axis, as we will now  show. 

A necessary condition for the crossing of the branch cut is $\Im [\lambda_k (\omega)] = 0$. Assume this holds for an eigenvalue $\lambdat_k(\omega)$ of $-\left (\Gh^R(\omega)\right )^{-1}$. Consider the eigenvalue equation ($\omega$-dependence suppressed) $ \left ( \Sigmah^R + \hh - \omega - i\eta \right ) | \lambdat_k \rangle = \lambdat_k | \lambdat_k \rangle$, giving 
\begin{align}
 \Im [\langle \lambdat_k | \Sigmah^R | \lambdat_k \rangle] - \eta = 0
\end{align}
We write $\Sigmah^R(\omega) = \hat{\Lambda}(\omega) -\pi i  \hat{\Gamma}(\omega)$, where $\hat{\Lambda}(\omega), \hat{\Gamma}(\omega)$ are Hermitian and the rate operator $\hat{\Gamma}(\omega)$ is PSD. We get
\begin{align}
 -2\pi \langle \lambdat_k | \hat{\Gamma} | \lambdat_k \rangle - \eta = 0. \label{nobranchcut}
\end{align}
Since $\langle \lambdat_k | \hat{\Gamma} | \lambdat_k \rangle \geq 0$, \Eq{nobranchcut} can not be satisfied. Thus, for a PSD approximation, no eigenvalue cross the real axis, and \Eq{hopeNoBranchCut} is valid.

\Eq{generalizedLuttinger} is a sum rule valid for those approximations that give analytic $\Gh^M(\zeta)$ and $\Sigmah^M(\zeta)$. Furthermore, if the approximation comes from a gauge-invariant $\Phi$, $I_2 = 0$. For example, the PSD approximations $GW$ and $GW_0$ will fulfill \Eq{generalizedLuttinger} with $I_2 = 0$, while the PSD approximation $G_0 W_0$ fulfills \Eq{generalizedLuttinger} with $I_2 \neq 0$ in general. We now give four different examples and limits of the sum rule. 

\subsection{Luttinger-Ward theorem}
In the case of the homogeneous electron gas, \Eq{generalizedLuttinger} leads immediately to what is known as the Luttinger-Ward theorem.~\cite{Luttinger1960a} See Ref.~\cite{Giuliani2005} for a more detailed discussion of this important sum rule in the electron gas. For homogeneous systems, the Hamiltonian and self-energy are diagonal in the momentum-basis, and $\Sigma^{R} (\bp,\mu) = \Sigma^A(\bp,\mu)$. Assuming a number-conserving approximation, $I_2 = 0$ and using 
\begin{align}
 \ln (x+i \eta) - \ln (x-i \eta) = 2 i \pi \theta(-x),
\end{align}
where $\theta$ is the step function, we obtain the Luttinger-Ward theorem
\begin{align}
 N = 2 V \int \frac{d \bp }{(2\pi)^3}
    \theta \parenth{\mu - \epsilon_\bp -\Sigma^R(\bp, \mu)},
\end{align}
where $\epsilon_\bp$ are the single-particle energies, $V$ is the volume and the factor of 2 accounts for spin. Thus, the Luttinger-Ward theorem is valid for those number-conserving approximations which retain the correct analytical properties.

\subsection{The Friedel sum rule}
\Eq{generalizedLuttinger} also allows for a useful relation in a completely different context, namely the Friedel sum rule in quantum transport. We consider a small region in space (e.g. a molecular region), where it is important to consider interaction effects carefully, coupled to macroscopic reservoirs (e.g. metallic leads) where a mean-field description is satisfactory. This implies that the frequency-dependent part of the self-energy is non-zero only in the central region. 

The sum rule of \Eq{generalizedLuttinger} still applies for the total number of particles, but usually we are more interested in the properties of the central region. The trace Tr$_C$ over the central region gives the number of particles $N_C$ in that region. By the use of embedding techniques \cite{Myohanen2008a,Cuevas2010} the Green's function in the central region can be written as
\begin{align}
\Gh^{R,A}(\omega) = \frac{1}{\omega - \hh_C - \Sigmah^{R,A}_{emb}(\omega) - \Sigmah^{R,A}(\omega) \pm i \eta},
\end{align}
where the embedding self-energy $\Sigmah_{emb}$ takes the environment into account in an exact way, and $\hh_C$ is the Hamiltonian for the  disconnected central region. The trace over the basis of the finite region gives the same type of sum rule as in \Eq{generalizedLuttinger},  provided we replace $\Sigmah$ by $\Sigmah + \Sigmah_{emb}$:
\begin{widetext}
\begin{align}
 N_C =
   \underbrace{\frac{1}{2\pi i}
   \traceC{ \ln \left [ \Sigmah^A(\mu) + \Sigmah^A_{emb} (\mu) + \hh_C - \mu + i\eta \right ] - \ln \left [ \Sigmah^R(\mu) + \Sigmah^R_{emb}(\mu) + \hh_C - \mu - i\eta \right ] }}_{I_1}
+  I_2. 
   \label{generalizedFriedel}
\end{align}
\end{widetext}

$\Sigmah^R_{emb}(\omega)$ enters $I_2$, which now has the form 
\begin{align}
   I_2 = -2 \Im  \left [ \int _{-\infty} ^{\mu} \frac{d\omega}{2\pi} \traceC{ 
   \Gh^R 
   \frac{\partial }{\partial \omega}  \left ( \Sigmah^R + \Sigmah^R_{emb} \right ) }  \right ].
   \label{genLuttingerIntegral}
\end{align}
We can thus split the contributions as $I_2 = I_{2,MB} + I_{2,emb}$. Now, $I_2 \neq 0$ in general, even for non-interacting systems. For a number conserving scheme, $I_{2,MB}=0$.

$I_1$ can be written in terms of the eigenvalues $\lambda_k(\omega)$ of $-\Gh^R(\omega)$. Since we will make use of this, we write out the form explicitly. 
\begin{align}
\begin{split}
 I_1 = \frac{1}{2\pi i}
   \traceC{ \ln \left [ -(\Gh^A(\mu))^{-1} \right ] - \ln \left [ -(\Gh^R(\mu))^{-1} \right ] } = \\
   =\frac{1}{2\pi i}
   \traceC{ \ln \left [ -\Gh^R(\mu) \right ] - \ln \left [ -\Gh^A(\mu) \right ] } = \\
   =\frac{1}{\pi} \Im 
   \sum_k \ln \left [ \lambda_k(\mu) \right ].\label{FriedelEigenvalue}
\end{split}
\end{align}

To recover the usual formulation of the Friedel sum rule, we take our central system to consist of a single interacting site with energy $\epsilon$, and use the wide-band limit approximation, $\Sigma^{R/A}_{emb}(\omega) = \mp i \Gamma$. \Eq{generalizedFriedel} becomes
\begin{align*}
 N_C =
   \frac{1}{2\pi i} \left (
   \ln \left [ \SigmaRt + i\Gamma \right ] - \ln \left [ \SigmaRt - i\Gamma \right ] \right ) + I_{2,MB},
\end{align*}
where $\SigmaRt = \Sigma^R (\mu) + \epsilon - \mu$. Using 
\begin{align}
 \ln{(x + iy )} -\ln{(x - iy )} = 
  \pi  - 2\arctan \left ( \frac{x}{y}\right ),
\end{align}
we obtain 
\begin{align}
 N_C = \frac{1}{2} -  \frac{1}{\pi}\arctan \frac{ \Sigma^R(\mu) + \epsilon - \mu}{\Gamma } + I_{2,MB}.\label{FriedelSumRule}
\end{align}
By using a gauge independent $\Phi$, $I_{2,MB} = 0$, and we obtain the well-known Friedel sum rule\cite{Langreth1966,Langer1961}. Thus, we can regard \Eq{generalizedFriedel} as a generalized Friedel sum rule.

\subsection{Analytical properties in the Anderson model}\label{AndersonExample}

In this section, we show explicitly, for the single-impurity Anderson model in the zero-temperature limit, that even conserving approximations can have spectral functions that can become negative. The Anderson model is a simple model in quantum transport, which consists of a single interacting dot in contact to a featureless reservoir with coupling strength $\Gamma$. The Hamiltonian of the dot is 
\begin{align}
 \Hh = \epsilon \sum _\sigma \nh_\sigma + \frac{U}{2} \sum _{\sigma \sigma'} \chd_\sigma \chd_{\sigma'} \ch_{\sigma'} \ch_\sigma,
\end{align}
where $\nh_\sigma$ is the number operator for spin $\sigma$ and $U$ is the interaction strength. 
Furthermore, we consider non-magnetic situations, and as such $G^R_{\sigma \sigma'}(\omega) = \delta _{\sigma \sigma'} G^R(\omega)$. For the Anderson model,  $G^R(\omega)$ is 
\begin{align}
 G^R(\omega) = \frac{1}{\omega - \epsilon - \Sigma^R(\omega) + i\Gamma},
\end{align}
where $\Gamma$ is the coupling to the reservoir. 
As our approximation to $ \Sigma^R(\omega)$, we take various 2nd order diagrams; the single-shot 2nd Born approximation \Fig{fig:ringDiagrams}a) $+$ \Fig{fig:2ndOrderExchange}a), the single-shot 2nd order exchange approximation \Fig{fig:2ndOrderExchange}a), and the self-consistent 2nd order exchange approximation \Fig{fig:2ndOrderExchange}b). The 2nd Born approximation is PSD, while the two 2nd order exchange approximations are not. We write here the explicit expression for $\Sigmah(z_1,z_2)$ for a general system with local interactions, since we will make use of larger systems in Sec.~\ref{exampleGeneral}.
\begin{align}
\begin{split}
 \Sigma_{kl}(z_1,z_2) = \delta _{kl} \delta (z_1,z_2) U n_{k}(z_1) + \\
    + c \ U^2  G_{kl} (z_1,z_2)   G_{lk} (z_2,z_1) G_{kl} (z_1,z_2), \label{selfenergy2nd}
\end{split}
\end{align}
where the indices $k,l$ label the interacting sites. For the Anderson model, $k=l=1$. The first term is the Hartree-Fock contribution, with $n_k(z_1) = -i G_{kk}(z_1,z_1^+)$. The coefficient $c$ in \Eq{selfenergy2nd} is $c=-1$ for the 2nd order exchange diagram, \Fig{fig:2ndOrderExchange}, $c=2$ for the ring diagram, \Fig{fig:ringDiagrams}, and $c=1$ for 2nd Born. This simple structure is due to the interaction being space-local. Furthermore, since 2nd Born is a PSD approximation, we have immediately that the rate operator of 2nd order exchange is negative semi-definite. 

The explicit expression for $\Sigmah^R_{kl}(\omega)$ in equilibrium is obtained from the Langreth rules \cite{Langreth1966}, and Fourier transforming (for more details, see Ref.~\cite{Karlsson2014a}. )
\begin{align*}
 \Sigma _{kl} ^R (\omega) = &\delta _{kl} U n_k + c U^2 \iint \frac{d\omega' d\omega''}{(2\pi)^2} [ \\
  &G_{kl}^R (\omega') G_{lk}^<(\omega'') G_{kl}^> (\omega - \omega' + \omega'') + \\
  + &G_{kl}^< (\omega') G_{lk}^A(\omega'') G_{kl}^< (\omega - \omega' + \omega'') + \\
  + &G_{kl}^< (\omega') G_{lk}^<(\omega'') G_{kl}^R(\omega - \omega' + \omega'') ],
\end{align*}
where the lesser and greater Green's functions in equilibrium are given by the fluctuation-dissipation theorem\cite{stefanucci2013}
\begin{align}
G^<_{kl} (\omega) &= 2 \pi i f(\omega) A_{kl} (\omega) \\
   G^>_{kl} (\omega) &= -2 \pi i \left [ 1 - f(\omega) \right ] A_{kl} (\omega).
\end{align}

For single-shot 2nd order exchange, $\Sigma^R (\omega)$ was evaluated using a Green's function from a self-consistent Hartree-Fock calculation. The self-consistent 2nd order exchange calculations did not converge if we would start from the Hartree-Fock Green's functions, however (see below). Instead, as in  Ref.~\cite{white1992}, we performed self-consistency with smaller values of $U$, and then starting new calculations with this initial guess.

\begin{figure}
  \centering
   \includegraphics[width=0.48\textwidth]{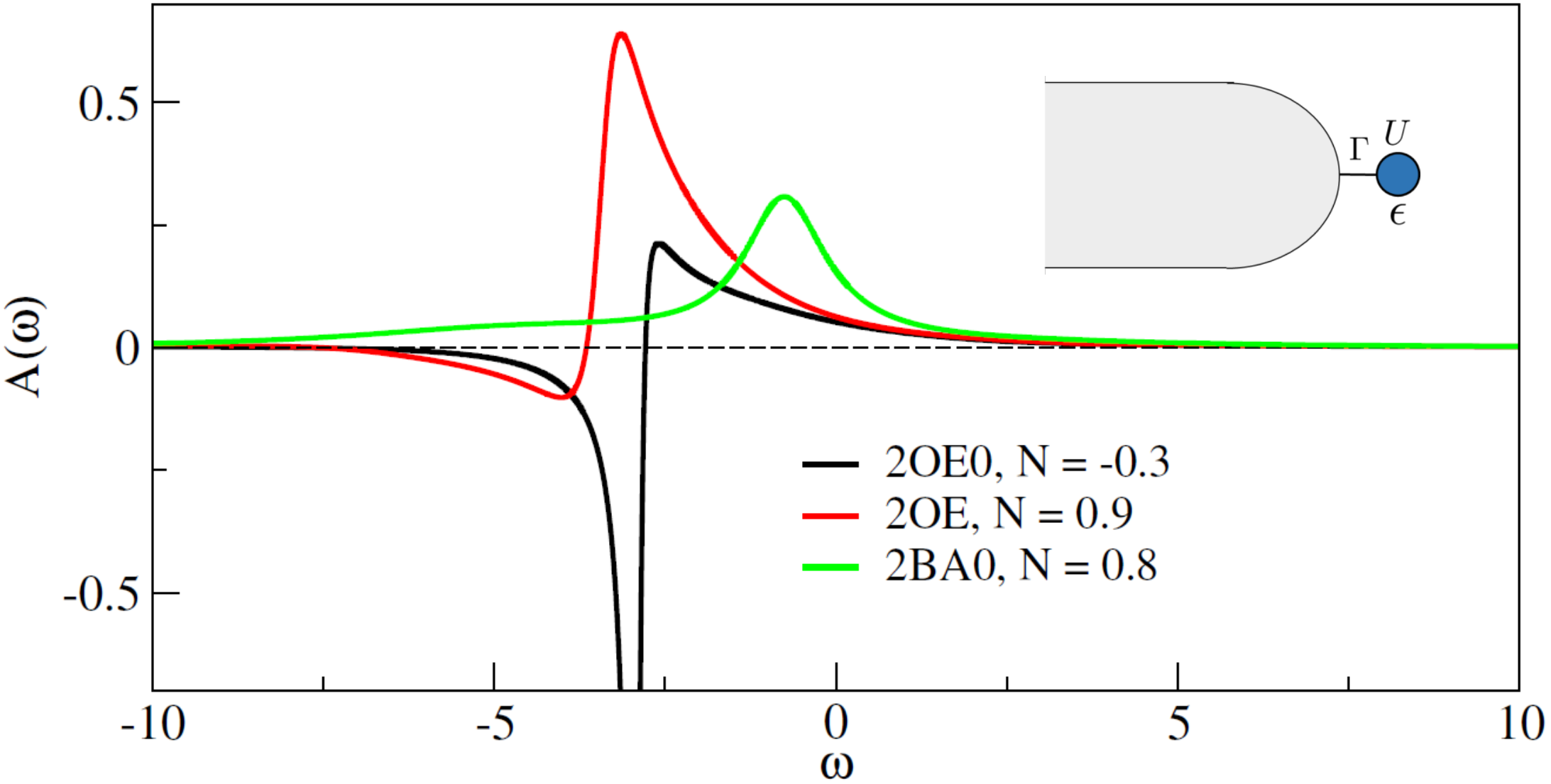}
  \caption{Spectral functions for the Anderson model (single impurity coupled to a featureless lead).
  2OE0: Single-shot 2nd order exchange diagram, \Fig{fig:2ndOrderExchange}a). 2OE: self-consistent 2nd order exchange, \Fig{fig:2ndOrderExchange}b). 2BA0: single-shot 2nd Born, \Fig{fig:ringDiagrams}a) $+$  \Fig{fig:2ndOrderExchange}a). 2OE0 and 2OE are non-PSD. 2nd Born, however, is PSD.   Compare with \Fig{fig:trackingAnderson}.
  The parameters are $U=6.5, \epsilon = -7$, and $\Gamma = 1$.   
   \label{fig:spectralFunc} }
 \end{figure}

In \Fig{fig:spectralFunc}, we show $A(\omega)$ for the three approximations. As expected, the 2nd order exchange approximation yields spectral functions which can become negative. In the one-shot case, one can by increasing $U$ make the spectral function negative in a large region. $\int d\omega A(\omega) = -0.2$ for one-shot 2nd order exchange, indicating a pole in the upper half-plane of $G^R(\zeta)$. For the other approximations, $\int d\omega A(\omega) = 1$. Similar to the case of $\alpha$ in Sec.\ref{simpleModel}, we find a crossover (not shown) with respect to $U$ above which a pole moves into the upper half-plane. Note also that since the particle number is $N_C= \int _{-\infty}^\mu A(\omega)$, it can become negative. However, the particle number defined via the spectral function can be unphysical due to the pole structure, see the discussion of the next section.

\begin{figure}
  \centering
   \includegraphics[width=0.48\textwidth]{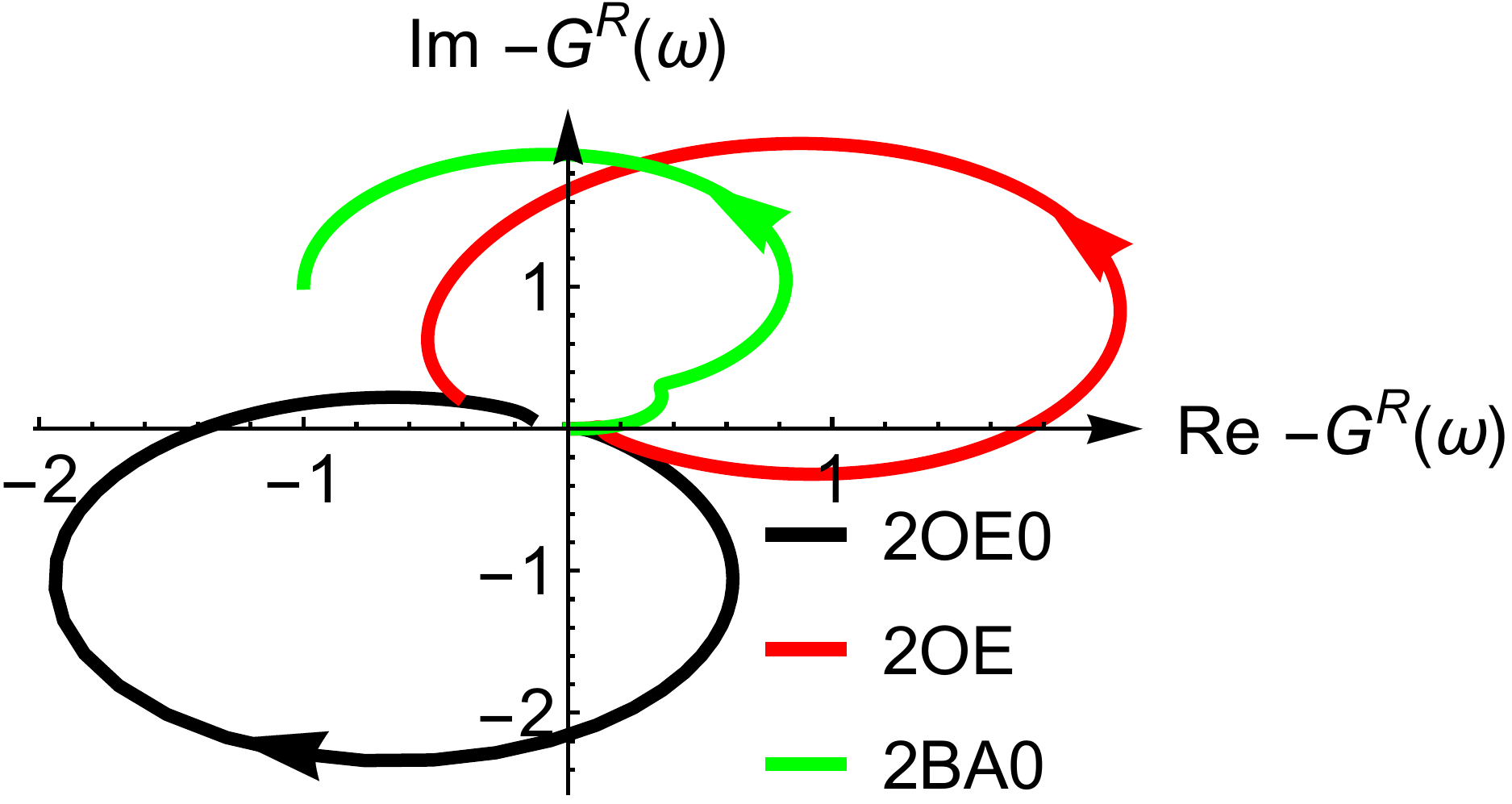}
  \caption{$-G^R(\omega)$ parametrized  with $\omega \in (-\infty,\mu]$, starting from $\omega=-\infty$ where  $G^R(-\infty) = 0$, and ending at $\mu = 0$, for the same situation as in \Fig{fig:spectralFunc}. The 2nd order exchange approximations are not PSD, and thus $-G^R(\omega)$ can cross the real axis. Single-shot 2OE breaks analyticity as well as the validity of the sum rules. Self-consistent 2OE does not. The single-shot 2nd Born approximation is PSD, and thus never crosses the real axis. For illustrative purposes, we have divided $G^R$ from 2OE0 with 3 and multiplied with 2 for 2BA0.
   \label{fig:trackingAnderson} }
 \end{figure}

We now discuss the Friedel sum rule, \Eq{generalizedFriedel}, for the Anderson model. The single-shot 2nd Born is PSD and thus fulfills the sum rule, but since the approximation is not particle conserving, $I_2 \neq 0$. The 2nd order exchange approximations are not PSD, and are thus not guaranteed to fulfill the sum rule. For our parameters, we found that self-consistent 2OE fulfilled the sum rule with $I_2 = 0$, while for single-shot 2OE, we find that $N_C - I_1 - I_2 = -2$.

To understand the term $-2$, we show in \Fig{fig:trackingAnderson} $-G^R(\omega)$ as parametrized by  $\omega$ from $\omega=-\infty$ to $\omega = \mu = 0$. As can be seen, $-G^R(\omega)$ from 2nd Born never crosses the real axis, which is guaranteed by its PSD property. However, $-G^R(\omega)$ obtained from single-shot 2OE crosses the negative real axis, and thus breaks the assumptions in the derivations of the  sum rules. Since we change Riemann sheet, but still use the principal branch of the logarithm, we miss an additional term of $2\pi i$. This translates into a term $-2$ in $I_1$, as can be seen from \Eq{FriedelEigenvalue}. The self-consistent 2OE also cross the real axis, but not at the branch cut. In fact, if we push the parameters further (increase $U$), analyticity breaks during the self-consistent cycle, and we are unable to reach convergence. This again underscores the importance of PSD approximations in self-consistent approximations.

Finally, we note that using only the second-order exchange diagram can seem like a constructed example. However, the same considerations apply if the  diagram is dressed with the screened interaction $W$. The second order diagram would then correspond to vertex corrections from $GW$. This approximation can yield a non-PSD spectral function \cite{Stefanucci2014}, and here we show that the same considerations apply in a model in quantum transport. 

\subsection{Example of the generalized sum rule}\label{exampleGeneral}
In the last example, we consider a 3-site linear chain at zero temperature to elucidate the generalized Friedel sum rule, \Eq{generalizedFriedel}, in a finite region. The Hamiltonian is 
\begin{align}
\Hh = \sum _{ij\sigma} h_{ij} \chd_{i\sigma} \ch_{j\sigma} + \frac{U}{2}\sum _{i\sigma \sigma'} \chd_{i\sigma} \chd_{i\sigma'} \ch_{i\sigma'} \ch_{i\sigma},
\end{align}
where $h_{ii} = \epsilon$, $h_{ij} = -1$ for nearest neighbors, and zero otherwise. As contacts, we choose two one-dimensional tight-binding leads with band-widths of $4$, attached to the first and third site, respectively, with coupling strength $-1$. The $3\times 3$ Green's function matrix for the central region is
\begin{align}
 \Gh^R(\omega) = \frac{1}{\omega - \hh_C - \Sigmah^R(\omega) - \Sigmah^R_{emb}(\omega) + i\eta},
\end{align}
where the embedding self-energy matrix $\Sigmah^R_{emb}(\omega)$ corresponding to the tight-binding leads, non-zero only for the first and third site, has an explicit expression, see Refs.~\cite{Myohanen2008a,Karlsson2014a,Cuevas2010}. 

We considered four different approximation to $\Sigmah^R(\omega)$, with various amounts of dressing: The self-consistent ring diagram \Fig{fig:ringDiagrams}d), the partially self-consistent ring diagram \Fig{fig:ringDiagrams}b), the single-shot ring diagram \Fig{fig:ringDiagrams}a), and the single-shot 2nd order exchange diagram \Fig{fig:2ndOrderExchange}. Of these, the first three approximations are PSD, and as such the total number of particles in the central region $N_C$ is given by the generalized Friedel sum rule, $N_C = I_1 + I_2$, where $I_1$ can be conveniently written in terms of the eigenvalues $\lambda_k(\omega)$ to $-\Gh^R(\omega)$, using \Eq{FriedelEigenvalue}, 
\begin{align}
 I_1 =
   \frac{1}{\pi}
   \sum_{k=1}^3 \ln \left [ \lambda_k(\mu) \right ],
\end{align}
and $I_2 = I_{2,MB} + I_{2,emb}$ is given by \Eq{genLuttingerIntegral}. As discussed before, for the fully self-consistent  and the  consistently dressed partially self-consistent approximations $I_{2,MB} = 0$, while for the single-shot cases  $I_{2,MB} \neq 0$, in general. To illustrate the sum rule, we also calculate $N_C$ according to $N_C = \int_{-\infty}^0 d\omega \trace{ \Ah(\omega)}$. 

\begin{figure}
  \centering
\includegraphics[width=0.48\textwidth]{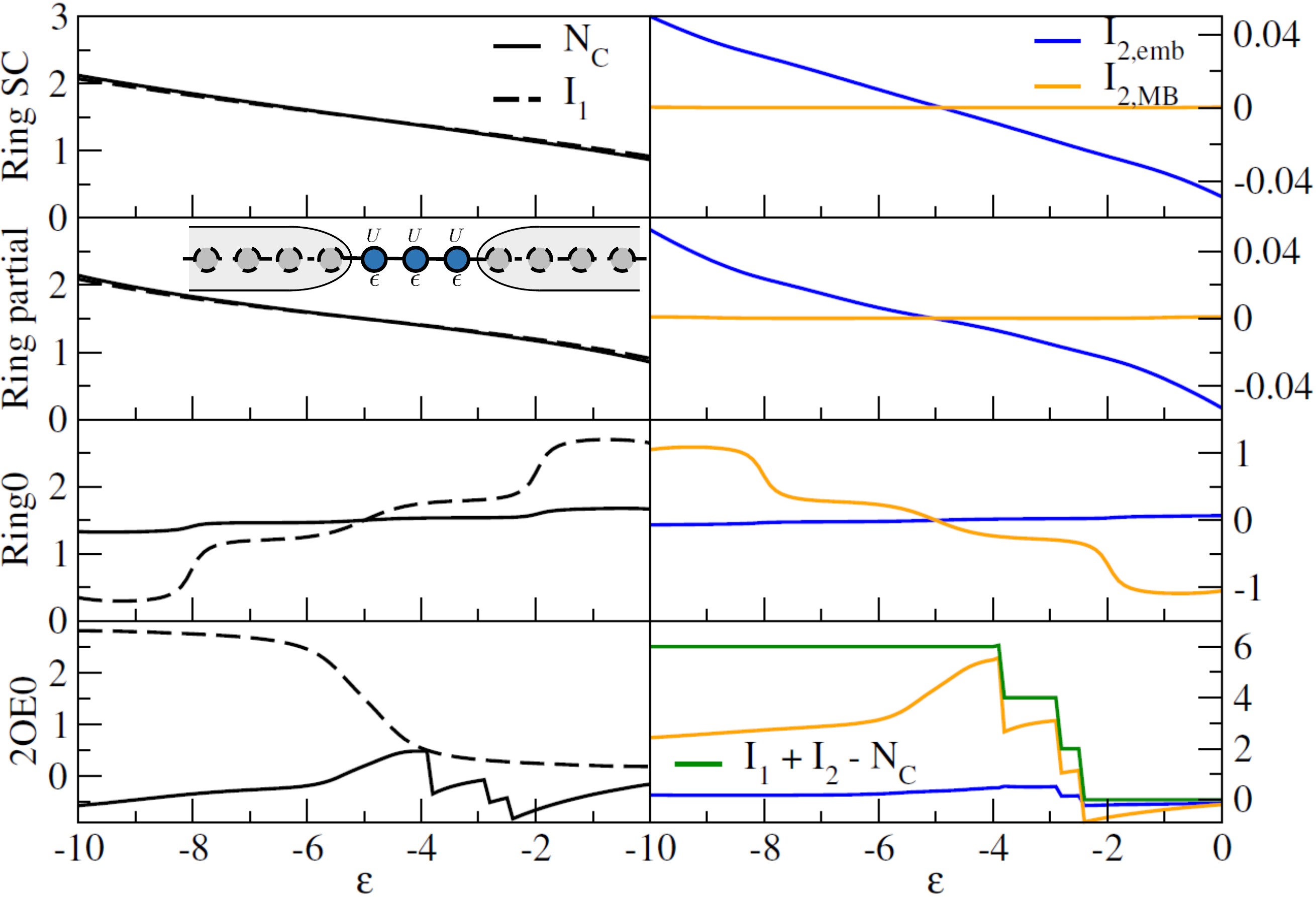}
  \caption{Number of particles $N_C$ in a 3-site system, for different dressings of ring approximations and single-shot 2nd order exchange. We show the self-consistent ring approximation (\Fig{fig:ringDiagrams}d)), partially dressed ring diagram, (\Fig{fig:ringDiagrams}b)), single-shot ring diagram (\Fig{fig:ringDiagrams}a)), and single-shot 2nd order exchange (\Fig{fig:2ndOrderExchange}a)).  $I_{2,emb}$ and $I_{2,MB}$ are the embedding and many-body contributions to the Luttinger integral, \Eq{genLuttingerIntegral}, respectively. Note the different scale for $I_2$ in the non-selfconsistent calculations.
   \label{fig:collect} }
 \end{figure}

To explore a large parameter range, we sweep with the gate voltage $\epsilon$, symmetrically from $\epsilon = -U/2$. $N_C$, $I_1$, $I_{2,emb}$ and $I_{2,MB}$ are shown in  FIG.~\ref{fig:collect}, for all the different approximations. We note that the correction from $I_{2,emb}$ is small, but becomes larger as we move away from $\epsilon = -U/2$. Note also that the contribution from $I_{2,emb}$ would increase with more sites connected to leads and lead coupling strength. We also see, as expected, that $I_{2,MB} = 0$ for the consistently dressed approximations. For the single-shot case, $I_{2,MB}$ can be very large. At $\epsilon = -U/2$, $I_{2,MB} = 0$, but is large away from this point. We plot $I_1$, which is the resulting particle number one would obtain if $I_2$ is not taken into account. The magnitude of $I_{2,MB}$ makes $N_C$ and $I_1$ very different, and in fact $I_1$ increases dramatically when raising the gate potential. Not taking $I_{2,MB}$ into account gave a similar erroneous trend for the Anderson model using single-shot 2nd Born in Ref.~\cite{Horvatic1980}, as was also pointed out by Ref.~\cite{white1992}. We also note that in the single-shot ring approximation, $N_C$ increases slightly as $\epsilon$ increases. This non-physical behavior is due to only taking the ring diagram into account. Single-shot 2nd Born (i.e. the  ring diagram and the 2nd order exchange diagram) does not exhibit this non-physical behavior (not shown). 

Finally, in the non-PSD 2nd order exchange approximation $N_C$ behaves non-physically and can even become negative. The norm of the spectral function (not shown) also becomes negative for certain parameter values, similarly to the case of the Anderson model. The change of norm indicates poles in the upper half-plane for $\Gh^R(\zeta)$, according to \Eq{normChange}. We also note that the particle number we calculate is $N_A = \int_{-\infty}^\mu d\omega \trace{ \Ah(\omega)}$. According to the discussion we had above, when there are poles in the upper half-plane for $\Gh^R(\zeta)$, with residue $a^R_k \neq 0$ and location $\xi^R_k$, $N_A$ does not have to equal $N=- i \trace{\Gh^M (\tau, \tau^+)} $, since we have to add the residues in \Eq{particleNumber2}. For simplicity, we discuss the situation in a single-site case.  From \Eq{normChange}, we have in the single-site case that 
\begin{align}
 1 = \int _{-\infty}^\infty d\omega A(\omega) + 2 \Re a^R. \label{normResidue}
\end{align}
At zero temperature, a pole can only contribute to the particle number if it is located to the left of $\mu$, 
 \begin{align}
  N = N_A + \theta(\mu-\Re \xi^R) 2 \Re a^R.
 \end{align}
Assuming that the jump in $N_A(\epsilon)$ occurs at the same time as when $\Re \xi^R(\epsilon)$ equals $\mu$, we found that the jump was canceled by the contribution from the residue, calculated from the norm in \Eq{normResidue}. This suggests that $N$ is a more physical quantity than $N_A$, in general. 

\begin{figure}
  \centering
\includegraphics[width=0.48\textwidth]{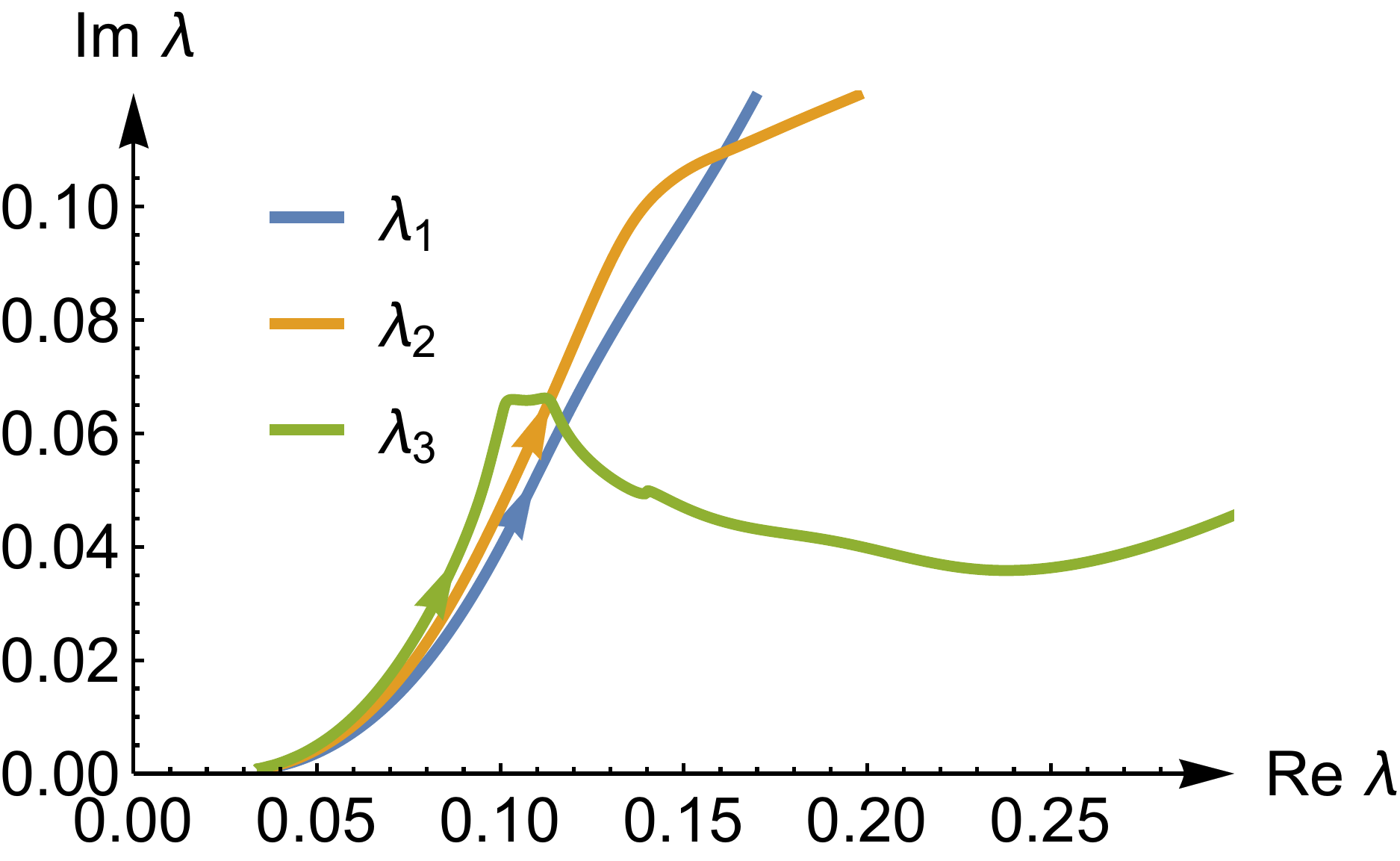}
  \caption{The three eigenvalues of $-\Gh^R(\omega)$ parametrized by $\omega$ for the system shown in \Fig{fig:collect}, for the fully self-consistent ring approximation, \Fig{fig:ringDiagrams}d). $\omega$ runs from $-\infty$ to $\omega = \mu = 0$. The parameters are $U=10$, $\epsilon = -U/2$. Being a PSD approximation, no eigenvalue can cross the real axis. Numerically we find $N_C - I_1 - I_2 = 0$.
   \label{fig:2ndorderRing} }
 \end{figure}

\begin{figure}
  \centering
\includegraphics[width=0.48\textwidth]{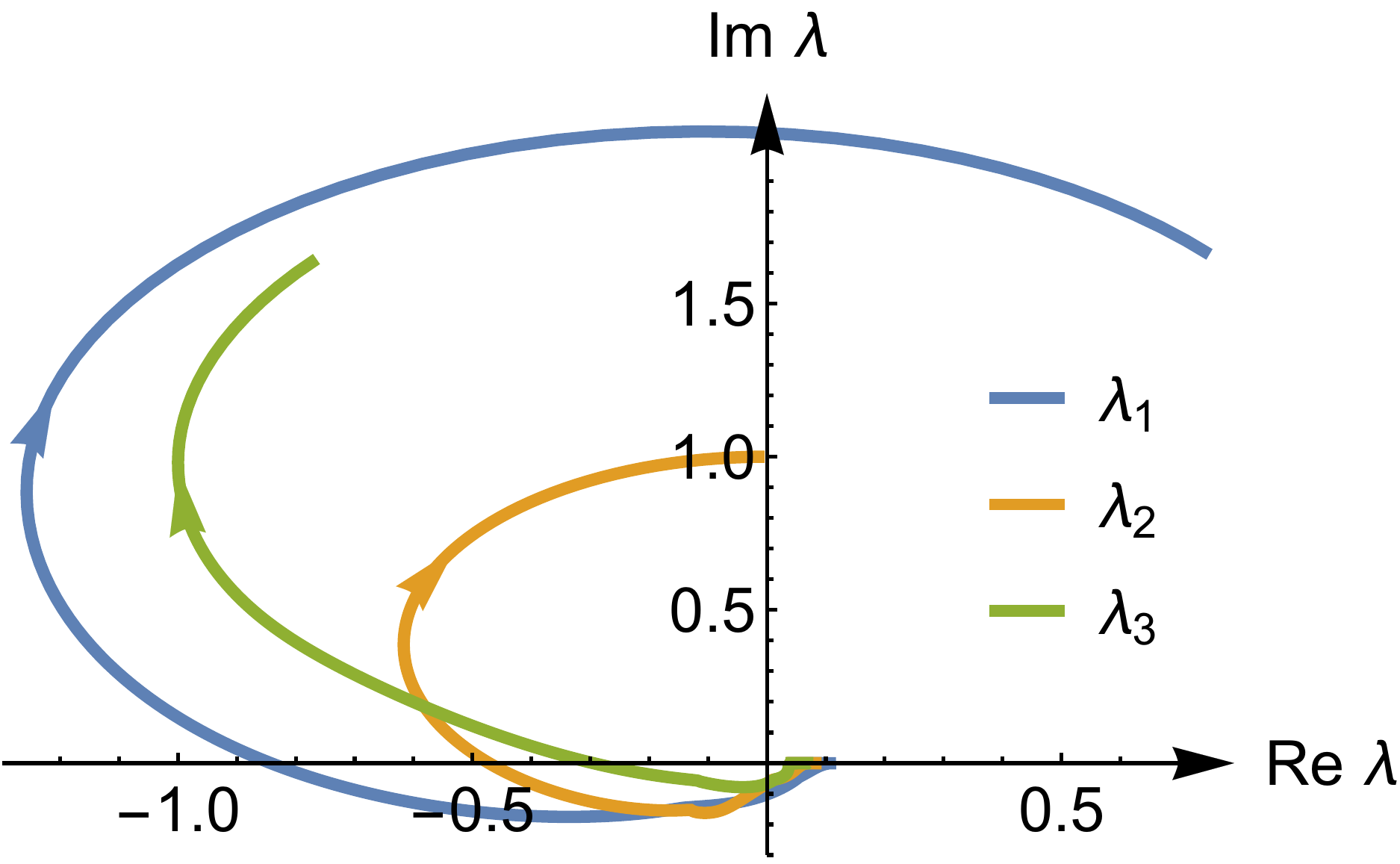}
  \caption{The three eigenvalues of $-\Gh^R(\omega)$ parametrized by $\omega$ for the system shown in \Fig{fig:collect}, for the single-shot 2nd order exchange, \Fig{fig:2ndOrderExchange}a). $\omega$ runs from $-\infty$ to $\omega = \mu = 0$.
  The parameters are $U=10$, $\epsilon = -U/2$. Being a non-PSD approximation, eigenvalues can, and do in this case, cross the real axis. Numerically, we find $N_C - I_1 - I_2 = -6$, since each eigenvalue yields a factor of $2$ to $I_1$.
   \label{fig:3x32OE0} }
 \end{figure}

Numerically, we find $N_C - I_1 - I_2 = 0$ to a high degree of accuracy for all the ring approximations. For the non-PSD 2nd order exchange approximation, we find instead that $N_C - I_1 - I_2$ varies, depending on the parameters. This can be understood by studying how the eigenvalues of $-\Gh^R(\omega)$ change as $\omega$ goes from $-\infty$ to $\mu$. To illustrate, we track $\lambda_k(\omega$ for the PSD self-consistent ring approximation, shown in \Fig{fig:2ndorderRing}, and the non-PSD single-shot 2nd order exchange approximation, shown in \Fig{fig:3x32OE0}. 

The eigenvalues $\lambda_k(\omega)$ of the PSD ring approximation in \Fig{fig:2ndorderRing} do not cross the real axis when we vary $\omega$, as expected. The eigenvalues of the non-PSD approximation in \Fig{fig:3x32OE0} do, however. For $\epsilon = -U/2$,  all three eigenvalues cross the branch cut. This means that, identical to the discussion we had above, for each crossing eigenvalue we obtain an additional factor of $2$ from $I_1$, which explains the result $N_C - I_1 - I_2 = -6$ in \Fig{fig:collect} for 2OE0. Depending on $\epsilon$, three, two, one, or no eigenvalues can cross the branch cut, the number of net crossings being $(I_1 +I_2 - N_C )/2$. As a side remark, we noted that eigenvalues could become degenerate for some frequencies (not shown). 

 
 \section{Conclusions}\label{conclusions}
We have extended the notion of $\Phi$-derivable schemes to include partially self-consistent and non-selfconsistent approximations. For such partially $\Phi$-derivable schemes. we have shown that by dressing the loops in $\Phi$ consistently, a gauge invariant approximation is obtained. This implies a fulfillment of the continuity equation at all times, which in turn implies number conservation. By making use of the non-equilibrium Green's function formalism, the resulting partially $\Phi$-derivable scheme can be applied in equilibrium as well as out of equilibrium, in finite and infinite systems. In equilibrium, partial $\Phi$-derivability allows for deriving a  generalized sum rule for the particle number, where we have stressed the importance of the correct analytical properties of the Green's functions. If the approximation is not gauge invariant, as is the case for many approximations used in the literature, we obtain another term $I_2$ in the sum rule. The rule was also applied to systems which allow for a partitioning into smaller subsystems, allowing for a generalized Friedel sum rule. The frequency-dependent embedding self-energies need to be taken into account in $I_2$. 

We have elucidated the known fact that PSD approximations automatically yield correct analytical properties for $G^R(\omega)$. The converse is not true, which we have exemplified in several systems; a spectral function that can become negative can still come from a $G^R(\omega)$ that has the correct analytical properties. We have proven that the validity of the Luttinger-Ward theorem and the Friedel sum rule depend crucially on the correct analytical properties of the Green's function. These considerations show that the simplest vertex corrections to $GW$ can lead to a violation of common sum rules. Extra diagrams need to be added to obtain a PSD spectral function, thus guaranteeing the satisfaction of the sum rules.

While a given approximation can be shown to be number conserving without making use of partial $\Phi$-derivability, we believe that this formalism is convenient and has several advantages. It is easy to see whether an approximation will be particle conserving or not, and also if the corresponding sum rules will be fulfilled. Combined with the knowledge of when the approximation has the correct analytical properties, these considerations could serve as a helpful tool in determining which diagrams to choose in a given situation. This could especially be important in, for example, diagrammatic Monte-Carlo integrations, where a large number of diagrams can be summed up.~\cite{Stefanucci2014,StefanucciErratum,Uimonen2015}

\begin{acknowledgments}
 D.K. and R.v.L. would like to thank the Academy of Finland for support under project no. 267839. 
\end{acknowledgments}

\appendix 
\section{} \label{AppendixA}
In this Appendix, we give more details about the zero-temperature limit of $\delta \Phi$,  \Eq{beforeWard}. We suppress the superscript $M$ for notational convenience.
Shifting the Matsubara Green's function  with $\delta \omega$, the variation is 
\begin{align}
 \delta \Gh(\omega_m) = \frac{\partial \Gh}{\partial \omega_m} \delta \omega
\end{align}
if $\delta \omega$ is small enough. From considering the spectral representation of $\Gh(\omega_m)$, \Eq{GMSpectral}, the requirement is that $ \delta \omega < |\omega_m|$. However, in relating $I_2$ to a variation in $\Phi$, we restricted the frequency shift to $\delta \omega = \frac{2\pi i }{\beta}$, meaning that $\omega_m + \delta \omega = \omega_{m+1}$. Thus, $\delta \omega$ is larger than the Matsubara frequencies $\omega_0 = \frac{\pi i}{\beta}$ and $\omega_{-1} = -\frac{\pi i}{\beta}$.

Moreover, for low temperatures, the variation $\delta \Gh(\omega_{-1}) = \Gh(\omega_0) - \Gh(\omega_{-1})$ yields a finite contribution as $T \to 0$, since $\omega_0$ and $\omega_{-1}$ are on different sides of the  branch cut at $\Im \zeta = 0$.  In fact, $\delta \Gh(\omega_{-1}) \to \lim _{\epsilon \to 0^+} \left [ \Gh(i\epsilon) - \Gh(-i\epsilon) \right ] = -2\pi i \Ah(\mu)$. 

To perform the zero-temperature limit, we thus separate the $m=-1$ and $m=0$ Matsubara points. 
\begin{align}
\begin{split}
 \sum _{m=-\infty} ^\infty e^{\eta \omega_m} \trace{\Sigmah (\omega_m) \delta \Gh(\omega_m) } = \\
 = \sum _{\substack{m=-\infty \\ m\neq -1, m \neq 0}} ^\infty e^{\eta \omega_m} \trace{\Sigmah (\omega_m) \frac{\partial \Gh}{\partial \omega_m} } \delta \omega + \\
 \!+ e^{\eta \omega_0} \Tr \! \left \{ \!\Sigmah  (\omega_0) \delta \Gh(\omega_0) \!\right \} \!\! +  \!
 e^{\eta \omega_{-1}} \Tr \! \left \{ \!\Sigmah  (\omega_{-1}) \delta \Gh(\omega_{-1}) \! \right \}\!.
 \end{split}
\end{align}
The sum is now well-defined in the $T \to 0$ limit, and we can use \Eq{zeroTemp}, yielding a principal-value integral. The other terms give $\delta \Gh(\omega_0) = \Gh(\omega_1) - \Gh(\omega_0) \to 0$ and  $\delta \Gh(\omega_{-1}) \to \lim_{\epsilon \to 0^+} \left ( \Gh(i\epsilon) - \Gh(-i\epsilon) \right )$. This gives
\begin{align}
\begin{split}
  \sum _{m=-\infty} ^\infty e^{\eta \omega_m} \trace{\Sigmah (\omega_m) \delta \Gh(\omega_m) } \to \\
  \to P \int _{-i\infty} ^{i\infty} \trace{\Sigmah (\zeta) \frac{\partial \Gh}{\partial \zeta} } d\zeta + \\
  + \trace{\Sigmah(0) \left [ \Gh(i\epsilon) - \Gh(-i\epsilon) \right ] } 
\end{split}
\end{align}
Thus, we have derived \Eq{beforeWard}.

\section{}\label{AppendixB}
We made use of several properties of matrix logarithms in the derivations of the sum rules, which will be more explained here. The key point is that matrix functions are not defined according to their Taylor expansions, since this assumes a radius of convergence. However, we cannot use a spectral representation, since Green's functions are not Hermitian, and as such not guaranteed to be diagonalizable.  

We assume that the Green's functions can be written as $L \times L$ matrices.  We also assume that the matrices are invertible, which they have to be in order to define a matrix logarithm. We make use of the fact that a general complex matrix can be written in its Jordan normal form, $\Qh = \Zh \Jh \Zh^{-1}$, where $\Jh$ is block diagonal with $M$ blocks $\Jh_i$ corresponding to non-degenerate eigenvalues. Each $m_i\times m_i$ block has the eigenvalue $\lambda_i$ of $\Qh$ on the diagonal, with multiplicity $m_i$. The superdiagonal of each block is equal to 1.
\begin{align}
 \Jh = \begin{bmatrix}
\Jh_1 & \;     & \; \\
\;  & \ddots & \; \\ 
\;  & \;     & \Jh_M\end{bmatrix}
\quad 
\Jh_i = 
\begin{bmatrix}
\lambda_i & 1            & \;     & \;  \\
\;        & \lambda_i    & \ddots & \;  \\
\;        & \;           & \ddots & 1   \\
\;        & \;           & \;     & \lambda_i
\end{bmatrix}.
\end{align}
Matrix functions can then be defined as (see, for example, Ref.\cite{Higham2008})
\begin{align}
 f(\Qh) = \Zh f(\Jh)\Zh^{-1} = \Zh \text{diag}(f(\Jh_i)) \Zh^{-1}
\end{align}
where 
\begin{align}
f(\Jh_i) = 
\begin{bmatrix}
f(\lambda_i) & f'(\lambda_i) & \cdots & \frac{f^{(m_i-1)} (\lambda_i)}{(m_i-1)!}   \\
\;        & f(\lambda_i)    & \ddots & \vdots  \\
\;        & \;           & \ddots & f'(\lambda_i)   \\
\;        & \;           & \;     & f(\lambda_i)       
\end{bmatrix} \label{matrixFunction}
\end{align}
This definition reduces to the usual spectral representation if $\Qh$ is diagonalizable. 
For matrix logarithms, \Eq{matrixFunction} can also be derived by writing $\Jh_i = \lambda_i(\one+\Kh)$, where $\Kh$ is lower triangular with 0 on the diagonal. $\ln \Jh_i = \ln (\lambda_i)\one + \ln (\one+\Kh)$ since all eigenvalues of $\Kh$ are real.~\cite{Higham2008} Furthermore, $\ln ( \one+\Kh) = \sum_{k=1}^{m_i-1} (-1)^{k+1}\frac{\Kh^k}{k}$ terminates, since $\Kh$ is strictly triangular and thus $\Kh^{m_i}=0$. 

\Eq{matrixFunction} immediately yields \Eq{traceLog},
\begin{align}
 \trace{\ln \left [ \Qh(\omega) \right] } = \sum_k \ln \left [ \lambda_k(\omega) \right ],
\end{align}
where  it is explicit that $\Qh$ must be invertible for the matrix logarithm to exist, since then all $\lambda_k(\omega) \neq 0$. 

If we furthermore assume that $\Qh(\omega)$ and the transformation matrices are smooth functions of $\omega$ we obtain \Eq{derivativeMatrix} as 
\begin{align}
\begin{split}
  \trace{\Qh^{-1}(\omega) \frac{\partial}{\partial \omega} \Qh(\omega)} = 
  \trace{J^{-1}(\omega) J' (\omega)} = \\
  = \sum _k  \frac{\lambda'_k(\omega)}{\lambda_k(\omega)} = 
  \frac{\partial}{\partial \omega} \trace{ \ln \left[ \Qh(\omega) \right ] }.
\end{split}
\end{align}

\begin{figure}
 \centering
 \includegraphics[width=0.42\textwidth]{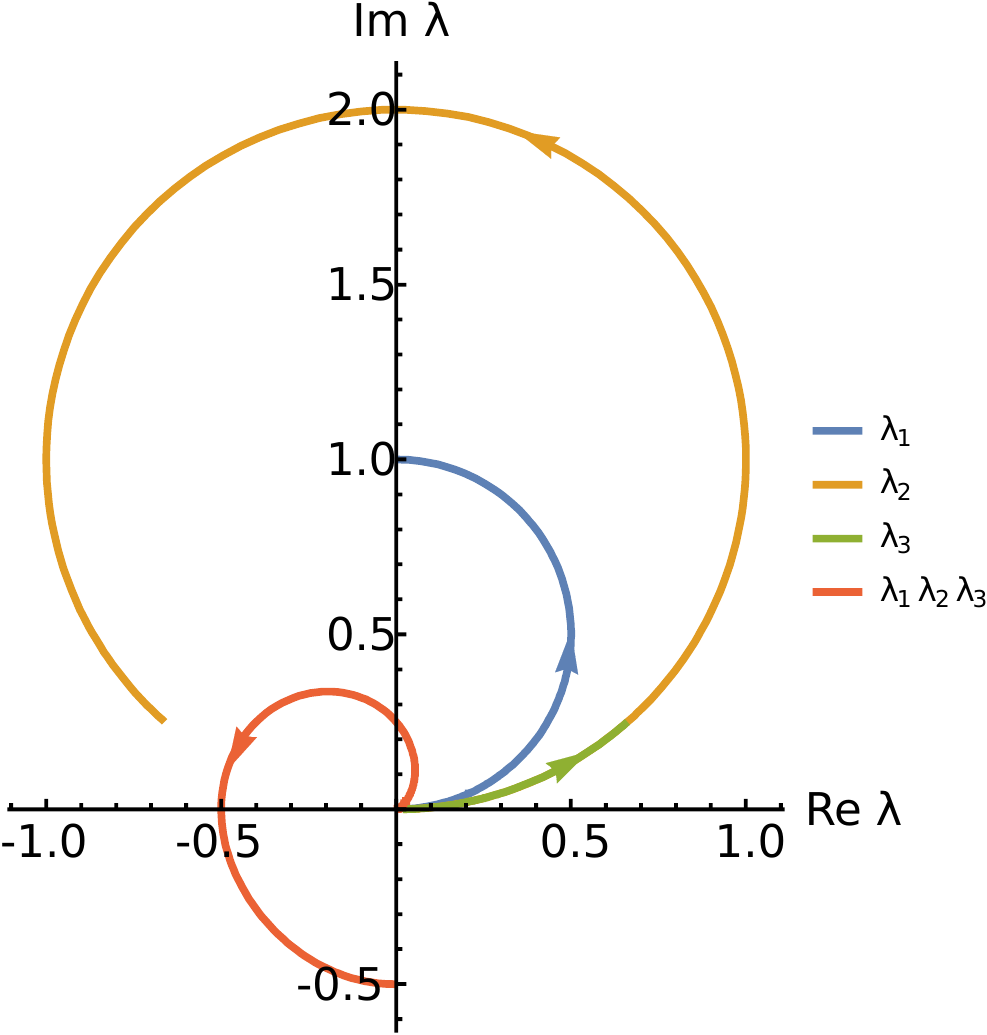}
 \caption{The eigenvalues $\lambda_k(\omega)$ of $-\Gh^R(\omega)$ for the 3-site system in Sec.\ref{exampleGeneral}, with $U=\epsilon = 0$ and using the wide-band limit for simplicity.  The curves starts at $\omega=-\infty$, for which $\lambda_i(-\infty) = 0$, and end at $\mu = 0$. Neither of the eigenvalues cross the real axis, while their multiplication does.}
 \label{fig:trackingEigenvaluesDeterminant2}
\end{figure}

These relations can be combined to derive \Eq{hopeNoBranchCut} as 
\begin{align}
\begin{split}
 \int _a ^b \frac{\partial}{\partial \omega} \trace{ \ln \left ( \Qh (\omega) \right )} d\omega = 
  \sum _k \int _a ^b  \frac{\lambda'_k(\omega)}{\lambda_k(\omega)}d\omega = \\
  = \sum _k \int _a ^b  d \ln \left [ \lambda_k (\omega) \right ].
\end{split}
\end{align}
The integrals on the right hand side are called logarithmic integrals.\cite{Gamelin2003} If the $\lambda_k(\omega)$:s do not cross the branch cut at $(-\infty,0)$, unique primitives can be found for each eigenvalue, yielding \Eq{hopeNoBranchCut}:
\begin{align}
\begin{split}
 \int _a ^b \! \! \frac{\partial}{\partial \omega} \trace{ \ln \left ( \Qh (\omega) \right )} \! d\omega \! = \! 
  \trace{\ln \left ( \Qh(b) \right )\! - \! \ln \left (\Qh(a) \right )}.
\end{split}
\end{align}
As a side remark, we note that we can write a logarithmic integral using determinants,
\begin{align}
\begin{split}
 \int _a ^b \trace{\Qh^{-1}(\omega) \frac{\partial}{\partial \omega} \Qh(\omega)} = 
 \int _a ^b \frac{\text{det} \left \{ \Qh(\omega)\right \}'}{\text{det}\left \{\Qh(\omega) \right \}} d\omega = \\
 = \int _a ^b d \ln \left [ \text{det} \left \{ \Qh(\omega) \right \} \right ] = 
 \int _a ^b d \ln \left [ \Pi_k \lambda_k (\omega) \right ].
 \label{determinant}
\end{split}
\end{align}
In this case, however, it is not possible to find a unique primitive in general, since the product of $\lambda_i(\omega)$ can cross the branch cut even if the individual $\lambda_i(\omega)$:s do not. This can be easily seen in the case of a non-interacting 3-site model. We consider the model in Sec.\ref{exampleGeneral}, and put $U=0, \epsilon=0$ and use the wide-band limit approximation. The eigenvalues of $-\Gh^R(\omega)$ are 
$\lambda_1 = -\frac{1}{\omega+i \Gamma}$ and \mbox{$\lambda_{2,3} = -\frac{2}{\left(2 \omega +i \Gamma \pm \sqrt{8-\Gamma ^2}  \right)}$}, neither of which cross the real axis. The multiplication of them $\lambda_1 \lambda_2 \lambda_3 = -\frac{1}{ \left ( \omega+i \Gamma \right ) \left (  \omega^2+i \Gamma  \omega - 2  \right ) }$, does cross the real axis, at $\omega=\pm 1$. Thus, if $\mu > -1$, we cannot find a unique primitive logarithm in \Eq{determinant}. The situation is depicted in FIG.\ref{fig:trackingEigenvaluesDeterminant2}.

\section{}\label{AppendixC}
In Sec.~\ref{PSD}, we made use of the spectral representation for $\Gh^M(\zeta)$, \Eq{GMSpectral}, and $\Sigmah^M(\zeta)$, \Eq{SigmaSpectral}. Using the Lehmann representation, one can show that the exact  $\Gh^M(\zeta)$ and $\Sigmah^M(\zeta)$ can be written in this form. In this Appendix, we will show that any analytic operator
with certain boundedness properties at infinity can be written in a spectral form. Furthermore, we will discuss how the spectral representation changes if the functions are non-analytic. We first consider scalar functions, and then generalize to operators. 

Suppose that we are given a function $f^M(\zeta)$, analytic except for single poles away from the real axis and decaying to zero at infinity, i.e. $f (\zeta) \to 0$ for
$|\zeta| \to \infty$. 
The poles have residue $a_k$ located at $\xi_k$ in the upper and lower half-plane. We define functions  $f^R(\zeta)$ and $f^A(\zeta)$ such that  
\begin{align}
 f^M(\zeta) =
 \begin{cases}
 f^R(\zeta) \quad \Im \zeta > 0   \\
 f^A(\zeta) \quad \Im \zeta < 0. \label{fm}
 \end{cases} 
\end{align}
Consider the integral along a closed half-circle in the upper half-plane. From the residue theorem, for $\Im \zeta >0$, we get
\begin{align}
 \frac{i}{2\pi} \oint d\xi \ \frac{f^R(\xi)}{\zeta-\xi} = f^R(\zeta) - \sum _k \frac{a^R_k}{\zeta-\xi_k^R}
\end{align}
where $\xi_k^R$ is the $k$th pole of $f^M(\zeta)$ located in the upper half-plane, with residue $a^R_k = \lim_{\xi \to \xi^R_k} (\xi - \xi^R_k) f(\xi)$. Since $f(\zeta)$ is bounded at infinity, the integral on the half-circle vanishes. Thus, we can write, for $\Im \zeta > 0$,
\begin{align}
 f^R(\zeta) = \frac{i}{2\pi} \int_{-\infty}^\infty d\omega \frac{f^R(\omega)}{\zeta-\omega} + \sum _k \frac{a^R_k}{\zeta-\xi^R_k}
\end{align}
where we have analytically continued $f^R(\omega) = \lim _{\eta \to 0} f^R(\omega + i\eta)$ in order to perform the integration on the real axis. 
In the same way, for $\zeta < 0$, 
\begin{align}
 f^A(\zeta) = -\frac{i}{2\pi} \int_{-\infty}^\infty d\omega \frac{f^A(\omega)}{\zeta-\omega} + \sum _k \frac{a^A_k}{\zeta-\xi^A_k},
\end{align}
where we also analytically continued $f^A(\zeta)$ to the real axis.
Consider now the function 
\begin{align}
 f^M(\zeta) = \frac{i}{2\pi} \int_{-\infty}^\infty d\omega \frac{f^R(\omega) - f^A(\omega)}{\zeta - \omega} + \\ 
 + \sum _k \frac{a^R_k}{\zeta-\xi^R_k} + \sum _k \frac{a^A_k}{\zeta-\xi^A_k}.
\end{align}
This function satisfies \Eq{fm}, which can be seen by considering $f^M(\zeta) - f^{R/A}(\zeta)$, for $\Im \zeta > 0$ and $\Im \zeta < 0$, respectively. For $\Im \zeta  > 0$, for example,
\begin{align}
 f^M(\zeta) - f^R(\zeta) = - \frac{i}{2\pi} \int_{-\infty}^\infty d\omega \frac{f^A(\omega) }{\zeta - \omega} + \sum _k \frac{a^A_k}{\zeta-\xi^A_k} = \\
 =\frac{i}{2\pi} \oint d\omega \frac{f^A(\omega) }{\zeta - \omega} + \sum _k \frac{a^A_k}{\zeta-\xi^A_k} = 0,
\end{align}
where we have closed the contour in the lower half-plane. 

Defining the spectral function $A(\omega) = \frac{i}{2\pi} \left ( f^R(\omega) - f^A(\omega) \right )$, we obtain 
\begin{align}
 f^M(\zeta) = \int_{-\infty}^\infty d\omega \frac{ A(\omega) }{\zeta - \omega} 
 + \sum _k \frac{a^R_k}{\zeta-\xi^R_k} + \sum _k \frac{a^A_k}{\zeta-\xi^A_k}.
\label{theSpectral}
 \end{align}
 If we add the restriction that $[f^M(\zeta^*)]^* = f^M (\zeta)$, we have that 
\begin{align}
\begin{split}
 \int_{-\infty}^\infty d\omega \frac{ A^* - A }{\zeta - \omega} 
 + \sum _k \frac{(a^R_k)^*}{\zeta- (\xi^R_k)^*} - \frac{a^R_k}{\zeta-\xi^R_k}  + \\
 + \sum _k \frac{ (a^A_k)^*}{\zeta-(\xi^A_k)^*} - \sum _k \frac{ a^A_k}{\zeta - \xi^A_k} = 0.
\end{split} \label{difference}
 \end{align}
By integrating around a small circle around a chosen $\xi^R_m$, which is in the upper half-plane, we obtain that 
\begin{align}
 2 \pi i a_m^R = \sum _k \oint_{\xi^R_m} d\zeta \frac{ (a^A_k)^*}{\zeta-(\xi^A_k)^*}.
 \end{align}
Since $a_m^R \neq 0$, we must have that $a_m^R = (a^A_m)^*$ and $\xi^R_m = (\xi^A_m)^*$. From \Eq{difference} it then immediately follows that $A(\omega) = A^*(\omega)$. We can then write $f^M(\zeta)$, \Eq{theSpectral}, as 
\begin{align}
 f^M(\zeta) = \int_{-\infty}^\infty d\omega \frac{ A(\omega) }{\zeta - \omega} 
 + \sum _k \frac{a^R_k}{\zeta-\xi^R_k} + \sum _k \frac{ (a^R_k)^*}{\zeta-(\xi^R_k)^*}.
\label{theSpectral2}
 \end{align}
By considering $f^{R/A}(\omega) = \lim_{\eta \to 0}f^M(\omega \pm i\eta)$ in \Eq{theSpectral2}, we find that $\mp \frac{1}{\pi} \Im f^{R/A}(\omega) = A(\omega)$. While $A(\omega)$ is real, it is not necessarily positive and can have any sign. 

From \Eq{theSpectral2} we can see the equivalence between functions analytic for $\Im \zeta \neq 0$ and functions that can be written as an integral over a spectral function. This gives justification to the use of the spectral representation for $\Gh^M(\zeta)$, \Eq{GMSpectral}, and $\Sigmah^M(\zeta)$, \Eq{SigmaSpectral}, for approximations that guarantee the correct analytical properties. The full justification follows when we consider operators below. When the spectral function is PSD, $-\Gh^R(\zeta)$ is a function that maps the upper half-plane to itself. This is a so-called Nevanlinna function. Such functions are considered in, e.g., \cite{Teschl2014}, where it is shown that such functions can be written in a spectral representation. However, the notion of spectral function can be generalized to measures, which means that also non-differentiable spectral functions, such as the ones consisting of delta functions, can be considered.

If $f^M(\zeta)$ is not analytic away from the real axis, but has simple poles, the spectral function can still be defined via the real-frequency retarded or advanced Green's function. We can thus regard \Eq{theSpectral2} as a generalized spectral representation. If we would define the particle number as $N_A = \int _{-\infty}^\mu A(\omega)$, we would miss the contribution from the poles if they exist, and different definitions of particle number would yield different results. 

We now generalize to operators $\Fh^M(\zeta)$, which we take to have finite dimension, and impose the condition $[\Fh^M(\zeta^*)]^\dagger = \Fh^M (\zeta)$. We apply the same considerations as above to each matrix element $F^M_{ij}(\zeta)$. Thus, \Eq{theSpectral2} holds for each matrix element,  yielding

\begin{align}
 F_{ij}^M(\zeta) = \int_{-\infty}^\infty d\omega \frac{ A_{ij}(\omega) }{\zeta - \omega} 
 + \sum _k \frac{a^R_{kij}}{\zeta-\xi^R_{kij}} + \sum _k \frac{ (a^R_{kji})^*}{\zeta-(\xi^R_{kji})^*},
\label{theSpectral3}
 \end{align}
where $\Ah^\dagger (\omega) = \Ah(\omega)$. 
Each matrix element can have their own set of poles. We now define 
\begin{align}
 \sum _k \frac{a^R_{kij}}{\zeta-\xi^R_{kij}} = 
 \sum _l \frac{\alpha _{lij}}{\zeta-\xi^R_{l}},
\end{align}
in which the sum over $l$ goes over all the poles of all matrix elements of $\Fh^M(\zeta)$, and we defined the residue matrix
\begin{align}
\left ( \hat{\alpha}_l \right )_{ij} = \alpha _{lij} = 
\begin{cases}
 a^R_{lij} \text{ if $l$ correspond to pole in $\Fh^M_{ij}$} \\
 0 \text{ otherwise.} 
\end{cases}
\end{align}
Using this rewriting, we obtain the spectral representation in operator form 
\begin{align}
\Fh^M (\zeta) \! = \! \int_{-\infty}^\infty \! \! \! d\omega \frac{\hat{A}(\omega)}{\zeta-\omega} \! + \!
\sum_l \left (   \frac{\hat{\alpha}_l}{\zeta - \xi_l^R} +
  \frac{\hat{\alpha}^\dagger_l}{\zeta - (\xi_l^R)^*} \right )\! .
\end{align}

We see, that even in the presence of poles, we can define the spectral function operator as 
$\Fh^R(\omega) - \Fh^A(\omega) = - 2 \pi i \Ah(\omega)$. We have thus shown that analytic operators, whose matrix elements decay to zero at infinity, can be written in a spectral representation. This discussion fully justifies the use of the spectral representation for $\Gh^M(\zeta)$, \Eq{GMSpectral}, and $\Sigmah^M(\zeta)$, \Eq{SigmaSpectral}, under the assumption of analyticity.

\section{}\label{AppendixD}
The particle number can be obtained from the diagonal of the Green's function, \Eq{N_particle}. In this Appendix, we will show that we can also obtain the particle number from the grand canonical potential $\Omega$ in equilibrium, and that both definitions agree for a partially $\Phi$-derivable approximation. For the exact case,~\cite{Luttinger1960a} as well as for any conserving approximation,~\cite{Baym1962} $\Omega$ is given by the Luttinger-Ward functional, which we write in the Klein form as
\begin{align}
 \beta \Omega[G] =  \Phi - \traceg{\Sigmah \Gh} - \traceg{\ln (-\Gh^{-1})}.  \label{LW}
\end{align}
when $G$ is the Green's function of the physical system at hand. Here,
\begin{align}
\traceg{\Sigmah \Gh} = \int _\gamma d1 d2 \ \Sigma(1,2) G(2,1^+) \\
\traceg{\Gh} = \int _\gamma d1 \ G(1,1^+)
\end{align}
and the contour is taken to be the Matsubara contour, $\int_\gamma d1 G(1,1^+) = -i\int_0^{\beta} d\tau \int d\xb \ G(\xb,\tau;\xb,\tau^+)$. Note that the assumption of correct analytical properties is already implicit in \Eq{LW} since the logarithm has to be single-valued (see also the discussion below \Eq{traceLog}).

A partially $\Phi$-derivable scheme which is non-conserving will not have unambiguous total energies, since different energy functionals give different values.~\cite{PuigvonFriesen2010} The Luttinger-Ward functional, however, is variational with respect to $G$, meaning that $\frac{\delta \Omega}{\delta G} = 0$ when evaluated at a $G$ coming from the Dyson equation from a conserving approximation. Thus, partially $\Phi$-derivable schemes can give results close to the fully conserving scheme. \cite{stefanucci2013,Luttinger1960a,Caruso2013,Aryasetiawan2002,Dahlen2004,Dahlen2006,Stan2006,Stan2009a} The Luttinger-Ward functional has recently been evaluated at finite temperature in extended systems~\cite{Welden2016}.

Following Baym,~\cite{Baym1962} we vary \Eq{LW} with respect to the chemical potential. For a partially $\Phi$-derivable approximation, we have that 
\begin{align}
 \frac{\partial \Phi}{\partial \mu} = \traceg{\Sigmah \frac{\partial \Gh}{\partial \mu}}.
\end{align}
 Note that we do not allow $G_0$ in $\Phi[G,G_0]$ to depend on $\mu$. \Eq{LW} becomes 
\begin{align}
 \beta\frac{\partial \Omega}{\partial \mu} = 
 -\traceg{ \frac{\partial \Sigmah}{\partial \mu} \Gh} -
 \traceg{\Gh \frac{\partial \Gh^{-1}}{\partial \mu} }. \label{someStep}
\end{align}
From $\Gh^{-1}$ in Matsubara form, we obtain 
\begin{align}
 \Gh^{-1}(\omega_m) = 
  (\omega_m + \mu)\oneh - \hh - \Sigmah (\omega_m)
\end{align}
which gives 
\begin{align}
 \frac{\partial \Gh^{-1}}{\partial \mu} = 
 \oneh  - \frac{\partial \Sigmah}{\partial \mu}.
\end{align}
\Eq{someStep} simplifies, and we get
\begin{align}
 \beta\frac{\partial \Omega}{\partial \mu} =  - \traceg{G } =
 i\int_0^{\beta} d\tau \int d\xb \ G(\xb,\tau;\xb,\tau^+),
\end{align}
which gives the desired relation
\begin{align}
 \frac{\partial \Omega}{\partial \mu} = - N. \label{omegaPot}
\end{align}
Note that the derivation of \Eq{omegaPot} only makes use of partial $\Phi$-derivability, and that the assumption of gauge invariance is not needed.

\section*{References}

%
\end{document}